\documentclass[reviewcopy]{elsarticle}

\usepackage[reviewcopy]{adndt}
\usepackage{l ongtable}

\usepackage{amsmath}
\usepackage{lineno}
\usepackage{amssymb}

\usepackage{color}
\usepackage[usenames,dvipsnames]{xcolor}

\biboptions{square,sort&compress}
\bibpunct[]{[}{]}{,}{n}{}{;}
\begin{document}

\begin{frontmatter}

\journal{Atomic Data and Nuclear Data Tables}

%% Author, fill in article title here

\title{Hyperfine structures and Land\'e $g_J$-factors  for $n=2$ states in beryllium-, boron-, carbon-, and nitrogen-like ions 
from relativistic configuration interaction calculations}

%% Fill in author list here
  \author[ulb]{S. Verdebout}
  \author[ulb]{C. Naz\'e}
%  \ead{E-mail: user@some.where.net}
  \author[sweden]{P. J\"onsson\corref{cor1}}
  \ead{per.jonsson@mah.se}
  \author[lithuania]{P. Rynkun}
  \author[ulb]{M. Godefroid}
  \author[lithuania]{G. Gaigalas} 

  \cortext[cor1]{Corresponding author.}
%  \fntext[X]{First author footnote.}
%  \fntext[Y]{Second author footnote.}

  \address[ulb]{Chimie Quantique et Photophysique, CP160/09, Universit\'e Libre de Bruxelles, 
 B~1050 Brussels, Belgium}
  \address[sweden]{Faculty of Technology and Society, Group for Materials Science and Applied Mathematics, Malm\"{o} University, 205-06 Malm\"{o}, Sweden}
  \address[lithuania]{Institute of Theoretical Physics and Astronomy, Vilnius University, LT-01108 Vilnius, Lithuania}

\date{16.12.2002} %please do not use \today, use actual date of version

\begin{abstract}  
Energy levels, hyperfine interaction constants, and Land\'e $g_J$-factors are reported for $n=2$ states 
in beryllium-, boron-, carbon-, and nitrogen-like ions 
from relativistic configuration interaction calculations. Valence, core-valence, and core-core correlation effects are taken into account through single and double-excitations from multireference expansions to increasing sets of active orbitals. A systematic comparison of the calculated hyperfine interaction constants is made with values from the available literature.
\end{abstract}

\end{frontmatter}

%%% Keywords and subject classification are not used in ADNDT 
%\begin{keywords}
%Beryllium-like ions Boron-like ions Carbon-like ions Nitrogen-like ions Hyperfine structure Land\'e factor
%\end{keywords}

%%% The table of contents should start a new page. This command will
%%% automatically pull all the unstarred \section, \subsection and
%%% \subsubsection titles into the Contents. Starred versions need to be
%%% done manually using
%%%            \addcontentsline{toc}{[[sub]sub]section}{Section title}
%%% at the correct place. Examples are given below.

%%% The lists of data figures and data tables are created automatically
%%% by the \listofDfigures and \listofDtables commands.

\newpage

\tableofcontents
\listofDtables
\listofDfigures
\vskip5pc

%%%% Authors begin text of article here %%%

\section{Introduction}
The hyperfine interaction, i.e. the interaction between the electrons and the non-spherical part of the electromagnetic field produced by the nucleus, is important in a number of applications. The hyperfine interaction breaks the $J$ symmetry of the atom and may open forbidden transitions that can be used for diagnostic purposes
in plasmas \cite{Braetal:98a,Braetal:02a}. Combined with accurate measurements of hyperfine structure splittings, calculated atomic parameters can be used to extract nuclear quadrupole moments $Q$ \cite{SunOls:92a}. In a different setting the hyperfine interaction is a key factor in determining excited state nuclear $g$-factors from recoil-in-vacuum experiments \cite{Stoetal:2010a}. Hyperfine interaction is also important in astrophysics, high resolution solar, stellar spectra reveal isotope shifts and hyperfine structures of many spectral lines. To correctly interpret the spectra it is necessary to
include isotope shifts and hyperfine structures in a theoretical modeling of the line profiles~\cite{Kur:93a,APJsupp2014}.
Hyperfine interaction is also of theoretical interest and is a valuable and sensitive probe of both electron correlation and QED 
effects~\cite{Beietal:98a,Godetal:2001a,Kozetal:2010a}.

Strong magnetic fields have been detected in hot stars of types O, B, and A. Investigations of these magnetic fields require knowledge
of accurate Land\'e $g_J$-factors \cite{Bieetal:2010a}. There is also a need for data to support polarization measurements of the coronal magnetic field \cite{Braetal:2000a}.
Some experimental $g_J$-factors have been published in successive NIST compilations, but data for many levels are still lacking. 
From a more theoretical point of view Land\'e $g_J$-factors give valuable information about coupling conditions in atomic systems, and
they can also be used to identify and label states \cite{FroJon:2001a}.   
 
A number of computer codes have modules for computing hyperfine structures and Land\'e $g_J$-factors, e.g. \textsc{civ3}~\cite{Hib:75a}, 
\textsc{mchf}~\cite{Jonetal:93a}, 
\textsc{atsp2k}~\cite{Froetal:2007a}.
The former codes
are non-relativistic with relativistic corrections in the Breit-Pauli approximation. Hyperfine interaction is an inner property and is very sensitive to relativistic effects \cite{LinRos:74a}. For this reason the calculations in this work rely instead on the fully relativistic \textsc{grasp2k} code \cite{Jonetal:2007a,Jonetal:2013a}, which has modules both for
hyperfine structure~\cite{Jonetal:96c} and Land\'e $g_J$-factors~\cite{AndJon:2006a}. The purpose of the present work is to complement the data sets on the isotope shift electronic factors~\cite{Nazetal:2014a} along the beryllium, boron, carbon, and nitrogen isoelectronic sequences with hyperfine interaction constants and Land\'e g$_J$-factors. The values should also serve as reference for other theoretical work. 
\section{Computational procedure}
The \textsc{grasp2k} package is based on the well-established multiconfiguration Dirac-Hartree-Fock (MCDHF) method~\cite{Gra:2007a}. Here
 the atomic state functions (ASFs), describing the studied fine-structure states, are obtained as linear combinations of symmetry adapted configuration state functions (CSFs)
\begin{equation}
\label{ASF}
|\gamma PJM_J\rangle   = \sum_{j=1}^{N_{CSFs}} c_{j} |\gamma_{j}PJM_J \rangle.
\end{equation}
In the above expression
$P$, $J$, and $M_J$ are the parity and the angular quantum numbers, respectively. The $\gamma$ symbol denotes all the other appropriate labeling of the configuration state function, such as the orbital occupancy and the coupling scheme. The configuration state functions are built from products of one-electron Dirac orbitals
\begin{equation}\label{eq:rel_spin-orb}
|n \kappa m \rangle =\frac{1}{r} \left(
\begin{array}{c}
P(n\kappa;r) \chi_{\kappa m}(\Omega)\\
i Q(n\kappa;r) \chi_{-\kappa m}(\Omega)
\end{array}
\right)\;,
\end{equation}
where we introduce the single electron quantum numbers $n$ and $\kappa$.
Requiring the energy functional of the ASF with respect to the Dirac-Coulomb $N$-electron Hamiltonian
\begin{equation}
H_{\mbox{{\footnotesize DC}}} = \sum_{i=1}^N \left( c\, {\boldsymbol{\alpha }}_i \cdot {\boldsymbol{p}}_i + (\beta_i -1)c^2 + V^N_i \right) + \sum_{i<j}^N \frac{1}{r_{ij}},\label{Hdc}
\end{equation}
to be stationary
with respect to perturbations in the expansion coefficients $\{c_j\}$ in the multiconfiguration expansion leads to a matrix eigenvalue
problem. The stationary condition with respect to variations in the radial functions,
in turn, leads to a system of coupled integro-differential equations subject to boundary
conditions at the origin and the infinity~\cite{Gra:2007a}. In the relativistic self-consistent field (RSCF) procedure both these
problems are solved to self-consistency. Having a set of radial functions $\{P(n\kappa;r), Q(n\kappa;r)\}$, it is possible to find the expansion coefficients $\{c_j\}$ by solving the related matrix eigenvalue problem. At this relativistic configuration interaction (RCI) level, higher order terms can be added to the Dirac-Coulomb Hamiltonian~(\ref{Hdc}). Normally the Breit interaction operator
\begin{equation}
\label{eq:Breit}
         H_{\mbox{{\footnotesize Breit}}} =  - \sum_{i<j}^N \left[ \boldsymbol{\alpha}_{i} \cdot \boldsymbol{\alpha}_{j}\frac{ \cos(\omega_{ij} r_{ij}/c)}{r_{ij}}  \right.\\
         + \left. (\boldsymbol{\alpha}_{i} \cdot \boldsymbol{\nabla}_i ) (\boldsymbol{\alpha}_{j} \cdot \boldsymbol{\nabla}_j )\frac{ \cos(\omega_{ij}r_{ij}/c) -1}{\omega_{ij}^2 r_{ij}/c^2} \right]
\end{equation}
as well as leading quantum electrodynamic (QED) corrections, self-energy and vacuum polarization, are included~\cite{McKetal:80a}. 

Calculations can be done for single states, but also for portions of a spectrum in the extended optimal level (EOL) scheme, where optimization is on a weighted sum of energies~\cite{Dyaetal:89a}. Using the latter scheme a balanced description of a number of fine-structure states belonging to one or more configurations can be obtained in a single calculation. All calculations were performed with the new release~\cite{Jonetal:2013a} of the \textsc{grasp2k} code~\cite{Jonetal:2007a}.

\section{Computation of hyperfine structures and Land\'e $g_J$-factors}
Once the ASFs have been obtained, measurable properties such as hyperfine structures and Land\'e g$_J$-factors can be expressed in terms of reduced matrix elements of 
spherical tensor operators ${\bf T}^{(k)}$ of different rank $k$,

\begin{equation}
\langle \gamma~PJ || {\bf T}^{(k)} || \gamma'~P'J'\rangle\;.
\end{equation}
Inserting the CSF expansions, the matrix element above splits into a sum over reduced matrix elements between CSFs. Using Racah algebra techniques, these matrix elements, in turn, can be obtained as weighted sums over radial integrals where the weights are obtained from the angular integration~\cite{Gaietal:2001a}.

\subsection{Hyperfine structure}
The hyperfine interaction is due to the interaction between the electrons and the non-spherical part of the electromagnetic field produced by the nucleus. The corresponding contribution to the Hamiltonian can be written as a multipole expansion

\begin{eqnarray}\label{eq:hfs}
H_{\mbox{{\footnotesize hfs}}}=\sum_{k\ge1} {\bf T}^{(k)} \cdot {\bf M}^{(k)}\;,
\end{eqnarray}
where ${\bf T}^{(k)}$ and ${\bf M}^{(k)}$ are spherical tensor operators of rank $k$ in the electronic- and nuclear-spaces, respectively~\cite{Arm:71a}. The terms with $k=1$ and $k=2$ are, respectively, the magnetic dipole and the electric quadrupole parts of the hyperfine interaction. Explicit forms of the electronic operators can be found in \cite{Jonetal:96c}. Higher order terms are much smaller and can often be neglected. By looking at~(\ref{eq:hfs}), it comes naturally that this operator does not commute neither with the total angular momentum of the electrons {\bf J} nor with the total angular momentum of the nucleus {\bf I}, but with the total atomic angular momentum {\bf F}={\bf I}+{\bf J}. Denoting the nuclear wave function by $| \kappa \pi I M_I \rangle$, where $\pi$, $I$ and $M_I$ are, respectively, the parity, the total angular momentum and its projection, the coupled states are expressed as
\begin{equation}
|\kappa\gamma~\pi P IJ F M_F \rangle = \sum_{M_I,M_J} \langle I J M_I M_J | F M_F \rangle | \kappa \pi I M_I  \rangle |\gamma PJM_J \rangle\;.
\end{equation} 
In first-order perturbation theory
the fine-structure levels $J$ are split into closely spaced hyperfine levels $F$ in atoms with non-zero nuclear spin $I$ according to
\begin{equation}
\langle \kappa\gamma~\pi P IJ F M_F |  {\bf T}^{(1)} \cdot {\bf M}^{(1)} +  {\bf T}^{(2)} \cdot {\bf M}^{(2)} | \kappa\gamma~\pi P IJ F M_F \rangle \;,
\end{equation}
which corresponds to the diagonal hyperfine effect. 
Expressing the magnetic dipole ($k=1$) and the electric quadrupole ($k=2$) contributions in terms of the reduced electronic and nuclear matrix elements, we get
\begin{eqnarray}
\langle \kappa\gamma~\pi P IJ F M_F |  {\bf T}^{(1)} \cdot {\bf M}^{(1)} | \kappa\gamma~\pi P IJ F M_F \rangle &=& (-1)^{I+J+F} 
\left\{ 
\begin{array}{c c c}
I & J & F \\
J & I & 1
\end{array} \right\} \sqrt{2J+1} \sqrt{2I+1} \nonumber\\
& &  \langle \gamma~P J ||  {\bf T}^{(1)} ||   \gamma~P J \rangle \langle \kappa~\pi I ||  {\bf M}^{(1)} ||   \kappa~\pi I \rangle \\
\langle \kappa\gamma~\pi P IJ F M_F |  {\bf T}^{(2)} \cdot {\bf M}^{(2)} | \kappa\gamma~\pi P IJ F M_F \rangle &=& (-1)^{I+J+F} 
\left\{ 
\begin{array}{c c c}
I & J & F \\
J & I & 2
\end{array} \right\} \sqrt{2J+1} \sqrt{2I+1} \nonumber\\
& &  \langle \gamma~P J ||  {\bf T}^{(2)} ||   \gamma~P J \rangle \langle \kappa~\pi I ||  {\bf M}^{(2)} ||   \kappa~\pi I \rangle \;.
\end{eqnarray}
Using the conventional definition of the nuclear magnetic dipole moment $\mu_I$ and the nuclear electric quadrupole moment $Q$, we define hyperfine interaction constants  that are not depending on the $F$ quantum numbers 
\begin{eqnarray}
A_J &=& \frac{\mu_I}{I} \frac{1}{\sqrt{J(J+1)}} \; \langle \gamma~P J ||  {\bf T}^{(1)} ||   \gamma~P J \rangle \\
B_J &=& 2\,Q\sqrt{\frac{J(2J-1)}{(J+1)(2J+3)}} \;  \langle \gamma~P J ||  {\bf T}^{(2)} ||   \gamma~P J \rangle \;.
\end{eqnarray}
The hyperfine energies are then given by
\begin{equation}
E_{\mbox{{\footnotesize hfs}}}(\gamma J F) = \frac{1}{2}A_J\,C+B_J\frac{\frac{3}{4} C(C+1)-I(I+1)J(J+1)}{2I(2I-1)J(2J-1)}\;,
\end{equation}
where $C=F(F+1)-J(J+1)-I(I+1)$.

The hyperfine interaction constants were evaluated using the hyperfine module of \textsc{grasp2k} \cite{Jonetal:2013a}. Angular data needed for the integration were saved on computer disc to reduce computational time for the full isoelectronic sequences. To allow an easy use of the provided results we have set the nuclear parameters to 1, i.e. $I=\mu_I=Q=1$. The reader may simply multiply the tabulated values of $A_J(I/\mu_I)$ and $B_J/Q$ by the factors $\mu_I/I$ and $Q$, respectively, to get the hyperfine interaction constants for the specific isotope of interest. The nuclear quantities have been tabulated by Stone~\cite{Sto:2005a}.
\subsection{Land\'e $g_J$-factor}
This work also provides the Land\'e g$_J$-factors of the considered levels all along the isolectronic sequences. The Land\'e g$_J$-factor is given by
\begin{eqnarray}\label{eq:gj}
g_J &=& 
\frac{2}{\sqrt{J(J+1)}} \langle\gamma PJ |\vert
\sum_{j=1}^N \left[ -i \frac{\sqrt{2}}{2\alpha^2} \, r_j 
\left( \mbox{\boldmath $\alpha$}_j\, {\mathbf C}^{(1)}_j \right)^{(1)}+ \frac{g_s-2}{2} \beta_j 
 \mbox{\boldmath $\Sigma$}_j \right] \vert \vert \gamma PJ \rangle \;, 
\end{eqnarray}
where $i = \sqrt{-1}$ is the imaginary unit, $\mbox{\boldmath $\Sigma$}$ the relativistic spin-matrix, and $g_s = 2.00232$
the $g$-factor of the electron spin corrected for QED effects. The Land\'e g$_J$-factor
determines the splitting of magnetic sub-levels in an external magnetic field~\cite{CheChi:85a}. In addition
it gives valuable information about the coupling conditions in the system~\cite{FroJon:2001a}. In the case of a pure $LSJ$ coupling scheme, the Land\'e factor is equal to 1 for singlet terms ($S=0$), equal to 2.00232 for $S$ states ($L=0$) and equal to 1.5012 for terms with $L=S$. The development of the transformation between the $jj$ and $LSJ$ coupling schemes~\cite{Gaietal:2003a,Gaietal:2004a}, however, provides more complete information.

The Land\'e $g_J$-factors were evaluated using the hyperfine module of \textsc{grasp2k} \cite{Jonetal:2013a}. For a complete write-up of the module see~\cite{AndJon:2006a}. 

\section{Electron correlation effects on the hyperfine structure}
The hyperfine structure is sensitive to different electron correlation effects. These effects need to be analyzed in order to assess the accuracy of the computed hyperfine structure constants in this work. In the Dirac-Fock approximation the charge distribution within a closed subshell is spherically symmetric, and the contribution to the magnetic dipole and electric quadrupole
interaction constants is zero. The contribution from a closed $s$ subshell is also zero, since the spin-densities from the two electrons are equal and the spin directions opposite.
Polarization of closed subshells in the core, due to the Coulomb interaction with open subshells, may have large effects on the hyperfine interaction constants.
The Coulomb exchange interaction reduces the repulsion between core and valence electrons with the same spin orientation, pulling the core electron towards the valence subshell.
Especially spin polarization is important. If the two $s$-electrons in the same subshell have different spin-densities a magnetic hyperfine interaction, which in the non-relativistic limit is referred to as the contact interaction, is induced~\cite{Froetal:97b,Godetal:97b}. As inner $s$-electrons have high densities at the nucleus, a very small imbalance is sufficient to cause
a net interaction, which is comparable to that of an open valence subshell. In addition to the spin polarization there is the orbital polarization, which is the distortion of the charge distribution of the core. The orbital polarization leads to contributions to the parts of the magnetic hyperfine interaction that in the non-relativistic limit are referred to as the orbital- and spin-dipolar interactions~\cite{Froetal:97b}. In addition there is an important contribution also to the electric quadrupole interaction.

In the MCDHF approach polarization effects are accounted for by including, in the wave function expansion, CSFs obtained by single excitations from the Dirac-Hartree-Fock reference configuration followed by a change in spin- and orbital coupling~\cite{Froetal:97b,Godetal:97b}. The CSFs that describe polarization effects are however energetically unimportant, and a large energy optimized orbital basis is needed to converge the contributions~\cite{Godetal:97b}. Although not used in the present work, non-orthogonal orbitals, specifically targeted for describing
polarization effects, drastically improve convergence~\cite{Veretal:2010a}. In addition to being affected by polarization effects, hyperfine structures are know to be sensitive also to higher order-correlation effects.

The importance of correlation effects are enhanced in cases where there are strong cancellations or when the interaction is zero in the Hartree-Fock (HF) or Dirac-Hartree-Fock (DHF) approximation.
As an example, relevant for the present study, we may look at $1s^22s^22p~^2P$ in B-like ions. For these states there are large, but canceling, contributions from the spin-polarization of the $1s$ and $2s$ subshells. Using only energy optimized orbitals, the convergence of the magnetic contact interaction is slow and oscillatory~\cite{JonFro:93a}. The cancellation also makes the  contact interaction sensitive to higher-order correlation effects. Another very difficult case is $1s^22s^22p^2~^3P_1$ in C-like ions. Here the $A_J$ constant is close to zero in the DHF approximation due to an almost perfect cancellation between the large orbital- and spin-dipole interaction constants. The contact interaction is zero. The introduction of polarization and correlation effects induces a contact interaction and changes the degree of cancellation between the orbital- and spin-dipole interaction constants, making  calculations extremely challenging~\cite{JonFro:93a}.
Among the studied states in this work there are several with half-filled $2p$ subshells, 
yielding $B_J$ constants that are zero at the non-relativistic HF level. Since the full interaction contribution comes from correlation effects, these $B_J$ constants are also very challenging to compute. Comparatively small relativistic effects, coupling different $LS$ terms, may also be of importance~\cite{CarGod:2011a,CarGod:2014a}. 
Thus, even at the low-$Z$ limit, a comparison between non-relativistic calculations and relativistic calculations may not be meaningful.   
\section{Generation of configurations expansions}
MCDHF and RCI calculations can be performed for single levels, but also for portions of a spectrum.  In this work calculations were carried out by parity and configuration, that is, wave functions for all states belonging to a specific configuration were determined simultaneously in an EOL calculation~\cite{Dyaetal:89a}. The EOL scheme has shown to be surprisingly efficient, and a balanced description of a large number of fine-structure states, belonging to one or more configurations, 
can be obtained in a single calculation~\cite{Jonetal:2013d,Ekmetal:2014a}. 
The expansions for the ASFs were obtained using the active set method \cite{Sturetal:2007a}. 
Here CSFs of a specified parity and $J$-symmetry are generated by excitations from a number of reference configurations to a set of relativistic orbitals. By applying restrictions on the allowed excitations, different electron correlation effects can be targeted. To monitor the convergence of the calculated energies and the physical quantities of interest, the active sets are increased in a systematic way by progressively adding layers of correlation orbitals. The configuration expansions strategies for the different ions, outlined in our previous work on isotope shifts~\cite{Nazetal:2014a} and included in the subsections below
for the sake of completeness, were aimed at generating wave functions with high overall quality, yielding accurate values for a number of expectation values such as energies, transition rates, and isotope shifts. Thus no special efforts were made to target electron correlation effects close to the nucleus by using, for example,  separately optimized orbitals accounting specifically for polarization effects~\cite{Godetal:97b}.  
\subsection{Boron-, carbon-, and nitrogen-like systems} 
The starting point was a number of MCDHF calculations, where the CSF expansions for the states belonging to a configuration were obtained by single and double (SD) excitations from a multireference (MR) consisting of the set of configurations that can be formed by the same principal quantum numbers as the studied configuration,
to an increasing active set of orbitals. The MCDHF calculations were followed by RCI calculations including the Breit
interaction and leading QED. The CSF expansions for the RCI calculations were obtained by SD-excitations from  a larger MR, including the most 
important configurations outside the complex, to the largest active set of orbitals. The correlation models used for the boron-, carbon-, and nitrogen-like ions are summarized in Table~A. The reference configuration is shown on the left. Columns two and three display the MR for, respectively, the MCDHF and RCI expansions. The largest active set is shown in column four, where the number of orbitals for each $l$-angular symmetry is specified.
The last column gives the number of CSFs for the RCI calculations. The described computational strategy has previously been used by Rynkun \textit{et al.}~\cite{Rynetal:2012a,Jonetal:2011a,Rynetal:2013a} to compute energies and transition rates, and more details  can be found in these papers. 

\subsection{Beryllium-like systems}
The beryllium-like systems are a little different. For each parity, the wave functions for all
states belonging to the $1s^22s^2$ and $1s^22p^2$ even configurations, and to the $1s^22s2p$ odd configuration, were determined simultaneously. 
For the MCDHF calculations, the CSF expansions were obtained by merging CSFs generated by SD-excitations from the reference configurations to an increasing active set with CSFs  obtained by single, double, triple, and quadruple (SDTQ) excitations to a subset of the active set of orbitals. The largest active set consisted
of orbitals with principal quantum numbers $n\leq8$ and the subset was formed by orbitals with principal quantum numbers $n\leq 4$. The MCDHF calculations were followed by RCI calculations including the Breit
interaction and leading QED. The expansions for the RCI calculations were obtained by merging CSFs generated by SDTQ-excitations from the reference configurations to the largest active set with the restriction that there in each CSF is at least two orbitals with $n \leq 3$ with CSFs generated by
SDTQ-excitations from the reference configurations to the active set of orbitals with principal quantum numbers $n\leq4$. The RCI expansions included about 296~000 relativistic CSFs. 

\begin{table}
\caption{Generated CSF expansions for the MCDHF and RCI calculations for the boron-, carbon-, and nitrogen-like ions.}
\begin{tabular}{llllr} \hline
Configuration         & MR for MCDHF             & \multicolumn{1}{l}{MR for RCI}                  & \multicolumn{1}{c}{Active set}  & \multicolumn{1}{c}{N$_{CSFs}$ in RCI} \\ \hline
\multicolumn{5}{c}{boron-like} \\ \hline
$1s^22s^22p$   & $1s^2\{2s^22p,2p^3\}$   & $1s^2\{2s^22p,2p^3,2s2p3d,2p3d^2\}$             & $\{9s8p7d6f5g3h1i\}$ &   200~100 \\
$1s^22p^3$     & $1s^2\{2s^22p,2p^3\}$   & $1s^2\{2s^22p,2p^3,2s2p3d,2p3d^2\}$             & $\{9s8p7d6f5g3h1i\}$  &   360~100 \\
$1s^22s2p^2$     & $1s^22s2p^2$            & $1s^2\{2s2p^2,2p^23d,2s^23d,2s3d^2\}$           & $\{9s8p7d6f5g3h1i\}$  &   300~100 \\ \hline
\multicolumn{5}{c}{carbon-like} \\ \hline
$1s^22s^22p^2$ & $1s^2\{2s^22p^2,2p^4\}$ & $1s^2\{2s^22p^2,2p^4,2s2p^23d,2s^23d^2\}$       & $\{8s7p6d5f4g2h\}$        &   340~100 \\
$1s^22p^4$     & $1s^2\{2s^22p^2,2p^4\}$ & $1s^2\{2s^22p^2,2p^4,2s2p^23d,2s^23d^2\}$       & $\{8s7p6d5f4g2h\}$         &   340~100 \\
$1s^22s2p^3$   & $1s^22s2p^3$            & $1s^2\{2s2p^3,2p^33d,2s^22p3d,2s2p3d^2\}$       & $\{8s7p6d5f4g2h\}$         & 1~000~100 \\ \hline
\multicolumn{5}{c}{nitrogen-like} \\ \hline
$1s^22s^22p^3$ & $1s^2\{2s^22p^3,2p^5\}$ & $1s^2\{2s^22p^3,2p^5,2s2p^33d,2s^22p3d^2\}$     & $\{8s7p6d5f4g1h\}$ &   698~100 \\
$1s^22p^5$     & $1s^2\{2s^22p^3,2p^5\}$ & $1s^2\{2s^22p^3,2p^5,2s2p^33d,2s^22p3d^2\}$     & $\{8s7p6d5f4g1h\}$ &   382~100 \\
$1s^22s2p^4$   & $1s^22s2p^4$            & $1s^2\{2s2p^4,2p^43d,2s^22p^23d,2s2p^23d^2\}$   & $\{8s7p6d5f4g1h\}$ &   680~100 \\ \hline
\end{tabular}
\end{table}

\newpage

\section{Results and discussion}
All the results of our calculations are presented in the tables~\ref{tab:HFS_Be},~\ref{tab:HFS_B},~\ref{tab:HFS_C}, and~\ref{tab:HFS_N} for beryllium-, boron-, carbon-, and nitrogen-like systems, respectively. 
In the following subsections, the results of the calculations are presented for the four isoelectronic sequences, and compared with  values from the  literature when available.
The identification of the levels is based on the $LS$ composition obtained by transforming from $jj$- to $LSJ$-coupling schemes using the {\sc jj2lsj} tool integrated in the new release of {\sc grasp2k}~\cite{Jonetal:2013a}. The two first CSFs with weight $|c_j|^2 \ge 0.1\%$ are also displayed in the level compositions.

\subsection{Beryllium-like systems}
Table~\ref{tab:HFS_Be} displays total energies, excitation energies, hyperfine magnetic dipole constants $A_J (I/\mu_I)$, electric quadrupole constants $B_J/Q$, and Land\'e $g_J$-factors. The reported isoelectronic sequence covers nuclear charges from $Z=5$ to $Z=74$. We investigate, for each nuclear charge, the six states $1s^22p^2~^1D_{2}$, $1s^22p^2~^3P_{1,2}$, $1s^22s2p~^1P^o_{1}$, and $1s^22s2p~^3P^o_{1,2}$, which are the lowest ones having hyperfine structures. One may note the interesting evolution along the sequence of the $LSJ$ composition of each state. At low $Z$ values, the states are quite pure and present almost no term mixing, but for high $Z$ values the spin-orbit interaction increases and term mixing appears.

 Table~\ref{tab:Litzen} presents the comparison between values from the present work and the data provided by Litz\'en \textit{et al.}~\cite{Litetal:98a} for B~II. This table brings to light the very good agreement that exists between these two theoretical works. The agreement is slightly better for the magnetic dipole constants than for the electric quadrupole constants. This is due to the more oscillating and slowly convergent behavior of the latter with respect to the increasing set of active orbitals~\cite{Veretal:2013a}.
Note that we had to scale the data provided by Litz\'en \textit{et al.}~\cite{Litetal:98a} ($I=1$, $\mu_I=1\mu_N$ and $Q=1$~barn) for allowing the comparison with our ``normalized'' hyperfine constants.

As quoted by Johnson~\cite{Joh:2010a}, Be-like  systems are the most studied systems beyond the He isoelectronic sequence for the study of hyperfine quenching. Systematic theoretical calculations of hyperfine interaction constants along this sequence have been performed by Marques \textit{et  al.}~\cite{Maretal:93a}, and Cheng \textit{et al.} \cite{Cheetal:2008a}.
Table~\ref{tab:Marques} compares our results with the ones obtained by Marques \textit{et  al.} \cite{Maretal:93a} and Cheng \textit{et al.} \cite{Cheetal:2008a}. The work by Marques \textit{et  al.} evaluates energies and wave functions  using the MCDHF method, including corrections for the Breit interaction and for the leading QED terms. As it has been pointed out by Cheng \textit{et al.} \cite{Cheetal:2008a}, some effects, as the correlation effects, are not enough considered, due to limited computational resources, in the pioneering calculations by Marques \textit{et  al.} \cite{Maretal:93a}. The work by Cheng \textit{et al.}  uses the RCI  method for evaluating energy levels and hyperfine matrix elements. We use the reported reduced matrix elements ($T(a,b)=\langle a||T^{(1)}||b\rangle$) associated with the magnetic dipole interaction, for computing the $A_J$ constants that are reported in table~\ref{tab:Marques}. The latter table shows that there is an almost perfect agreement between our values and the ones provided by Cheng \textit{et al.} \cite{Cheetal:2008a}. Due the very good agreement that Cheng \textit{et al.} \cite{Cheetal:2008a} gets with the experiment, we are therefore very confident in the values that we provide for this isoelectronic sequence.

\begin{figure}%[ht!]
\begin{minipage}[b!]{1\linewidth}
%\centering
\includegraphics[width=0.49\linewidth]{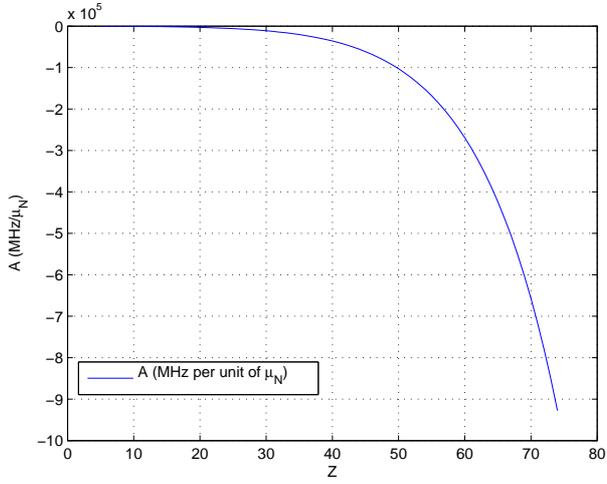}
\includegraphics[width=0.49\linewidth]{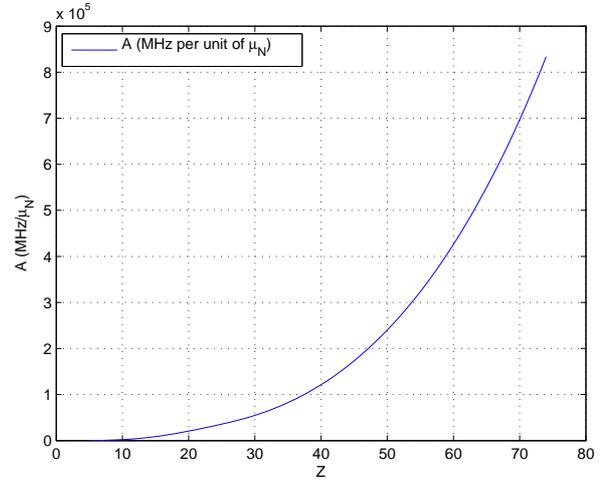}
\end{minipage}
\caption{Evolution along the beryllium isoelectronic sequence ($5\le Z \le 74$) of the hyperfine constants $A(I/\mu_I)$ for :\newline
(on left side) the third root characterized by  $J=1$. 
(on right side) the third root characterized by $J=2$. 
They present, respectively, a $1s^22p^2~^3\!P_1$ and a $1s^22p^2~^1\!D_2$ dominant character all along the sequence.
\label{fig:A_Be-like}}\end{figure}

\begin{figure}%[ht!]
\begin{minipage}[b!]{1\linewidth}
\centering
\includegraphics[width=0.49\linewidth]{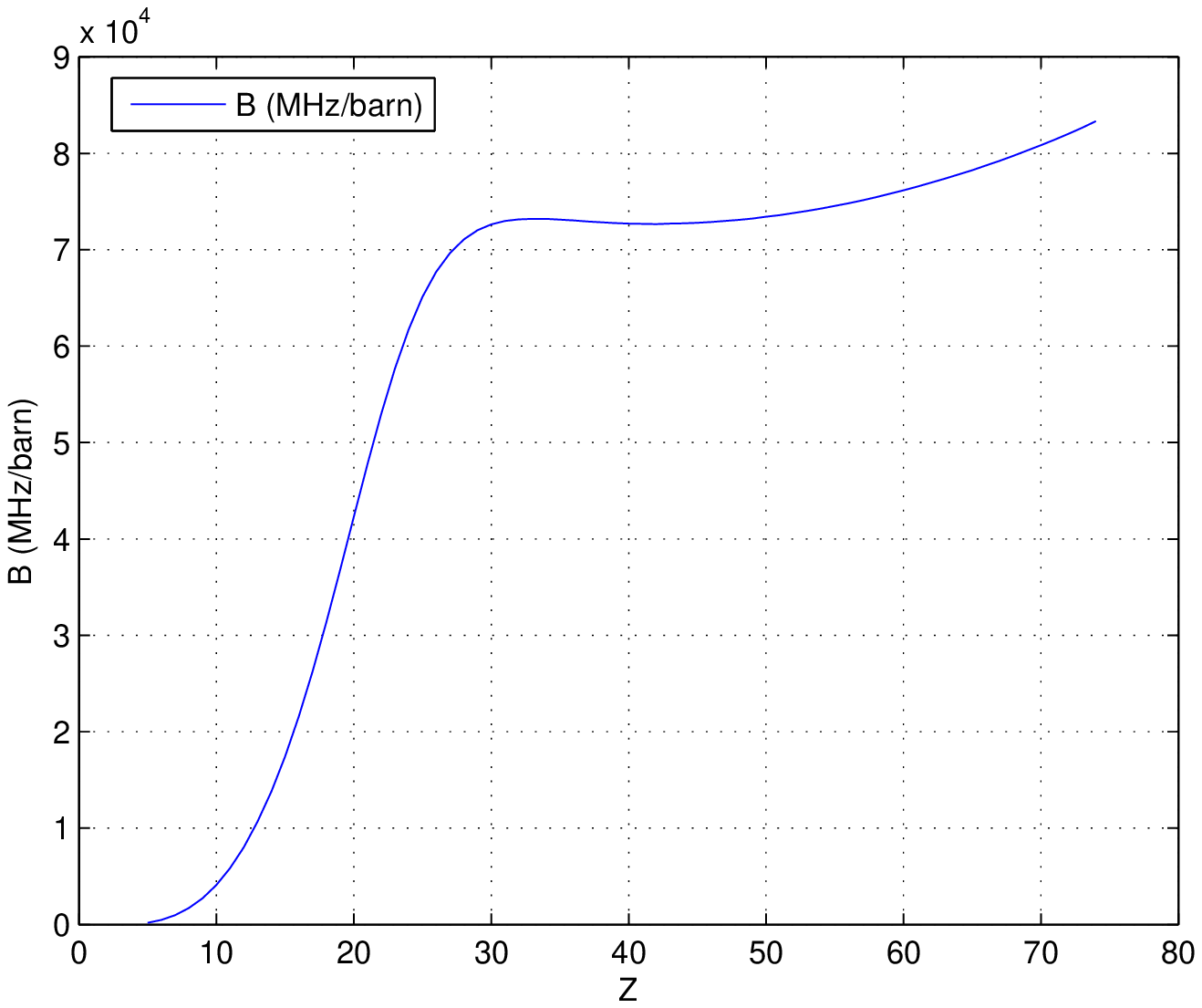}
\includegraphics[width=0.49\linewidth]{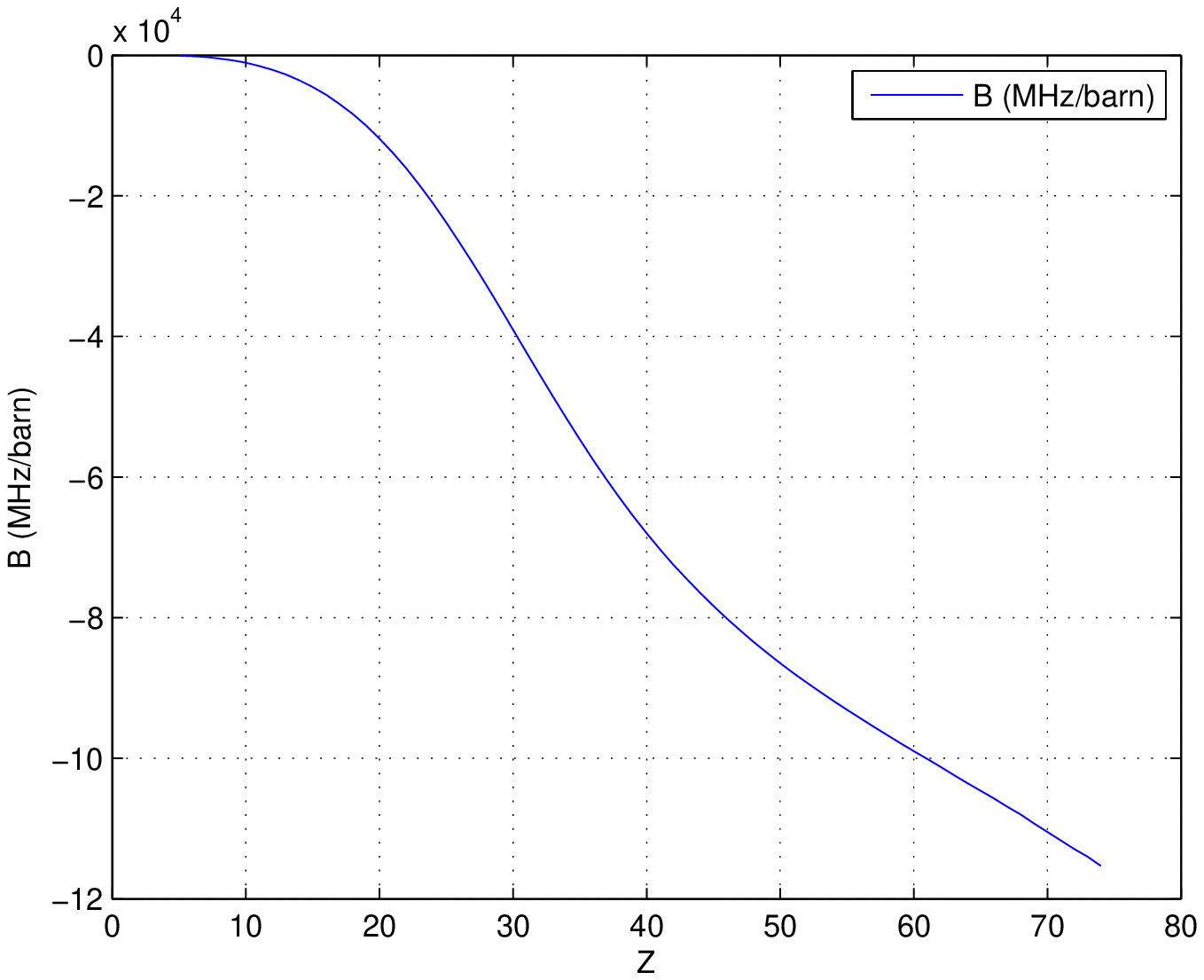}
\end{minipage}
\caption{Evolution along the beryllium isoelectronic sequence ($5\le Z \le 74$) of the hyperfine constants $B/Q$ for : \newline
(left side) the third root characterized by $J=2$.
(right side) the first root characterized by $J=1$. 
They present, respectively, a $1s^22p^2~^1\!D_2$ and a $1s^22s2p~^3\!P^o_1$ dominant character all along the sequence. 
\label{fig:B_Be-like}}\end{figure}

\begin{figure}
\begin{minipage}[b!]{1\textwidth}
\centering
\includegraphics[width=0.49\textwidth]{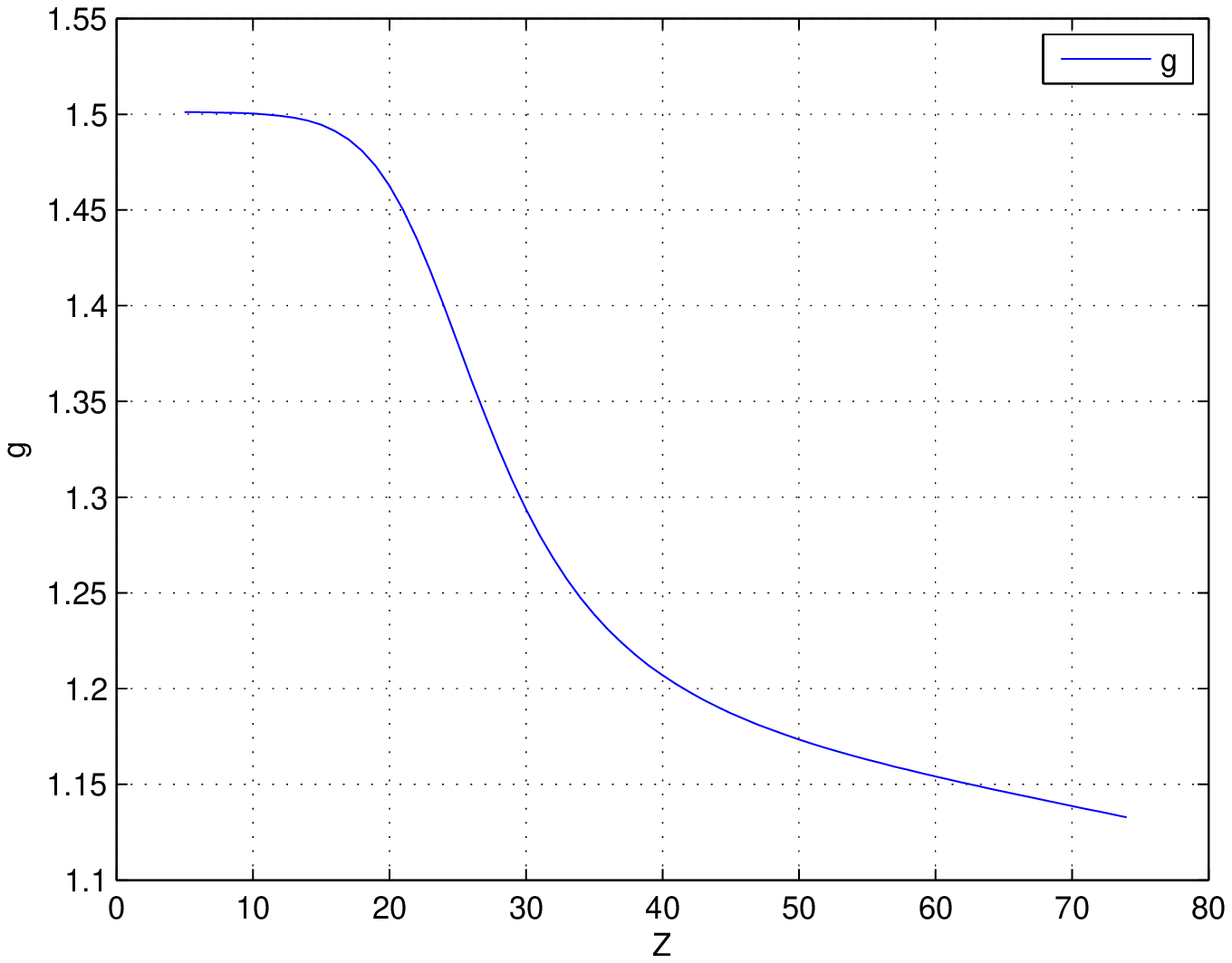}
\includegraphics[width=0.49\textwidth]{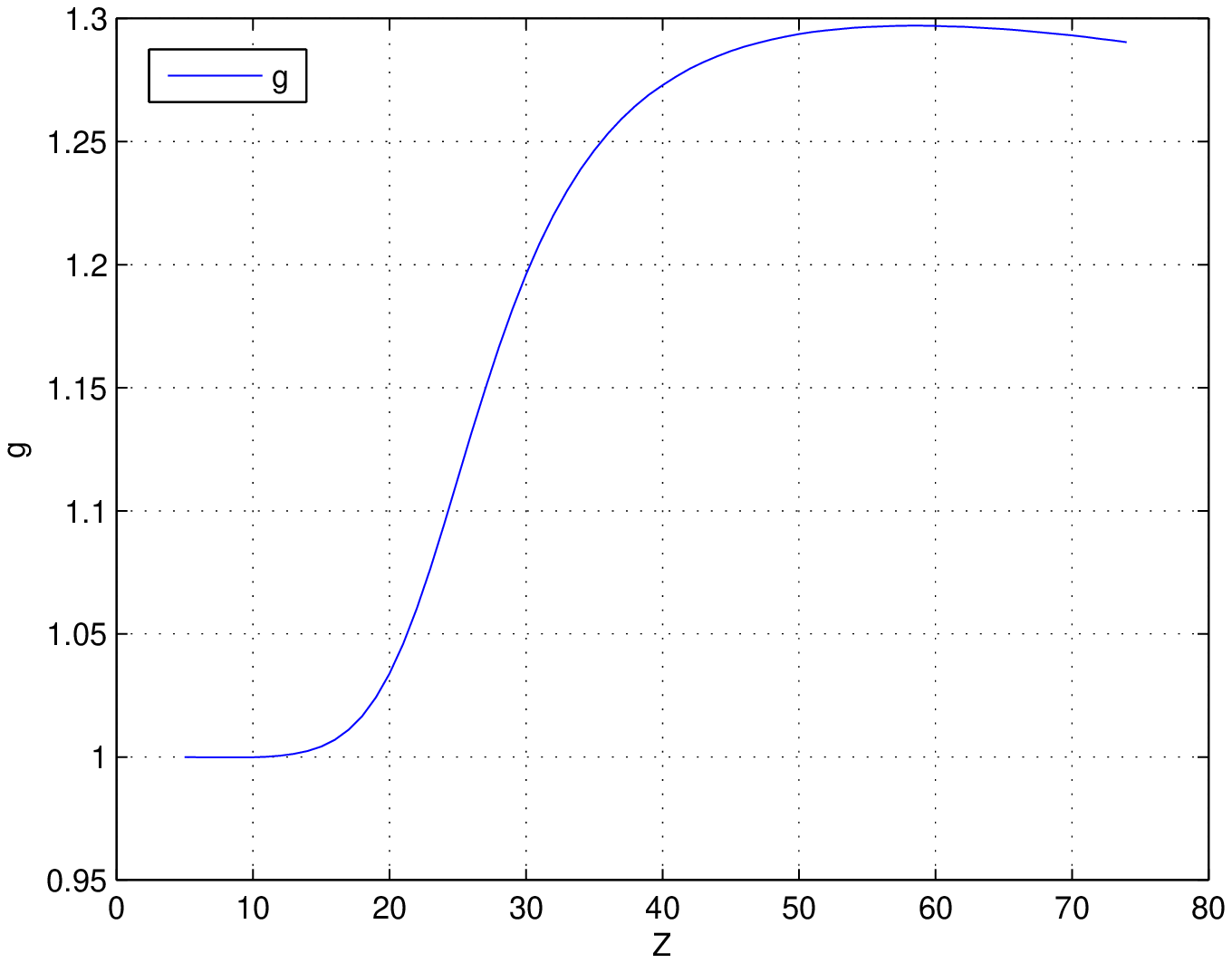}
\end{minipage}
\caption{Evolution along the beryllium isoelectronic sequence ($5\le Z \le 74$) of the Land\'e $g_J$-factor for : \newline
(left side) the second root characterized by $J=2$. (right side) the third root characterized by $J=2$.
They present, respectively, a $1s^22p^2~^3\!P_2$ and a $1s^22p^2~^1\!D_2$ dominant character all along the sequence. 
\label{fig:G_Be-like}}\end{figure}

It is often interesting to investigate atomic properties along an isoelectronic sequence to check their expected smoothness in the trend or possibly detect some anomaly in the calculation for a given ion, or to point out some problem in the level assignment. Figures~\ref{fig:A_Be-like}, \ref{fig:B_Be-like} and \ref{fig:G_Be-like} present, respectively, the evolution of $A(I/\mu_I)$, $B/Q$ and Land\'e $g_J$-factor along the beryllium isoelectronic sequence ($5 \le Z \le 74$) for a few chosen states. 
Other figures of isoelectronic evolution for all states are available on demand from the corresponding author. 

\subsection{Boron-like systems}
Table~\ref{tab:HFS_B} displays total energies, transition energies, hyperfine interaction constants, and 
Land\'e $g_J$-factors along the boron isoelectronic sequence. Once again, it is quite interesting to observe the rising of the spin-orbit interaction with increasing nuclear charge. The resulting term mixing becomes so important that for the molybdenum (Mo) ion, the dominant contributor, and then the dominant character, of two different states is the $1s^22p^3~^2P^o_{3/2}$ CSF. 
Recently, the O~IV ion has been studied by Dutta and Majumber~\cite{DutMaj:2012a} with the relativistic coupled-cluster method, J\"onsson \textit{et al.}~\cite{Jonetal:2010b}, with the multiconfiguration Dirac-Hartree-Fock method, and Sun \textit{et al.}~\cite{Sunetal:2012a} using Rayleigh-Ritz variation method with configuration interaction. Table \ref{tab:Dutta_Sun} compares the results of these calculations with the ones obtained in this work. 
For the $1s^22s^22p~^2P^o_{1/2,3/2}$ states, the multiconfiguration works agree very well, but differ
slightly with the values obtained by Dutta and Majumber. 
As for the states $1s^22s2p^2$ $^4P_{1/2,3/2,5/2}$, the electric quadrupole constants $B_J$ fit well with both Sun \textit{et al.}~\cite{Sunetal:2012a} and J\"onsson \textit{et al.}~\cite{Jonetal:2010b}. On the other hand, the present magnetic dipole $A_J$ values agree with the values of J\"onsson \textit{et al.}, but differ by around 2\% with the
values given by from  Sun \textit{et al.}.
Dutta and Majumber studied other ions along the B-like isoelectronic sequence up to $Z=21$. The third and fourth columns of table \ref{tab:Dutta} give the magnetic dipole constant $A_J(I/\mu_{I})$ in MHz per units of $\mu_N$, for the calculations of Dutta and Majumber and ours, respectively. Strangely enough, the values of the $^2P_{1/2}$ states agree within $\sim0.5\%$ while the values of the states $^2P_{3/2}$ agree only within $4\%$ or $5\%$.
To discriminate between the different calculations, and to further validate the present results, we performed large scale
multiconfiguration Hartree-Fock (MCHF) calculations for the states in O~IV. The MCHF calculations are fully variational and the expansions obtained by SD-excitations from large MR sets to an active set of orbitals with $nl \le9h$. 
The MCHF results were relativistically corrected by scaling the interaction constants with the ratio of DHF and HF values.
Results from the independent MCHF calculation are compared with the RCI values in table~\ref{tab:MCHF_OIV}. The two different calculations give values of $A_J$ that are in excellent agreement (to within less that 0.02 \% for the magnetic dipole constant). Based on this we can say that the present RCI calculations give very accurate values of the magnetic hyperfine interaction constants $A_J$ and, furthermore, that these values are more accurate than the values from the calculations by Dutta and Majumber ~\cite{DutMaj:2012a} and by Sun \textit{et al.}~\cite{Sunetal:2012a}.
Comparing the $B_J$ constants from the MCHF and RCI calculations we see that there are large differences for the $1s^22p^3~^2D_{3/2}$ and $1s^22p^3~^2P_{3/2}$
 states. The $B_J$ constants for these two states are zero at the HF level, making correlation and relativistic effects comparatively more important. For the two states above the  
difference between the $B_J$ constants from the RCI and MCHF calculations is due to the fact that the MCHF calculations do not include the $LS$ term mixing. The values from the RCI calculations are more accurate and the preferred ones.  
\subsection{Carbon-like systems}
Table~\ref{tab:HFS_C} presents the total and transition energies, the hyperfine interaction constants, the Land\'e $g_J$ factors, and the leading components of the many-electron wave function within the $LSJ$ coupling scheme. 

\begin{figure}[ht!]
\begin{minipage}[b!]{1\linewidth}
\centering
\includegraphics[width=0.32\linewidth]{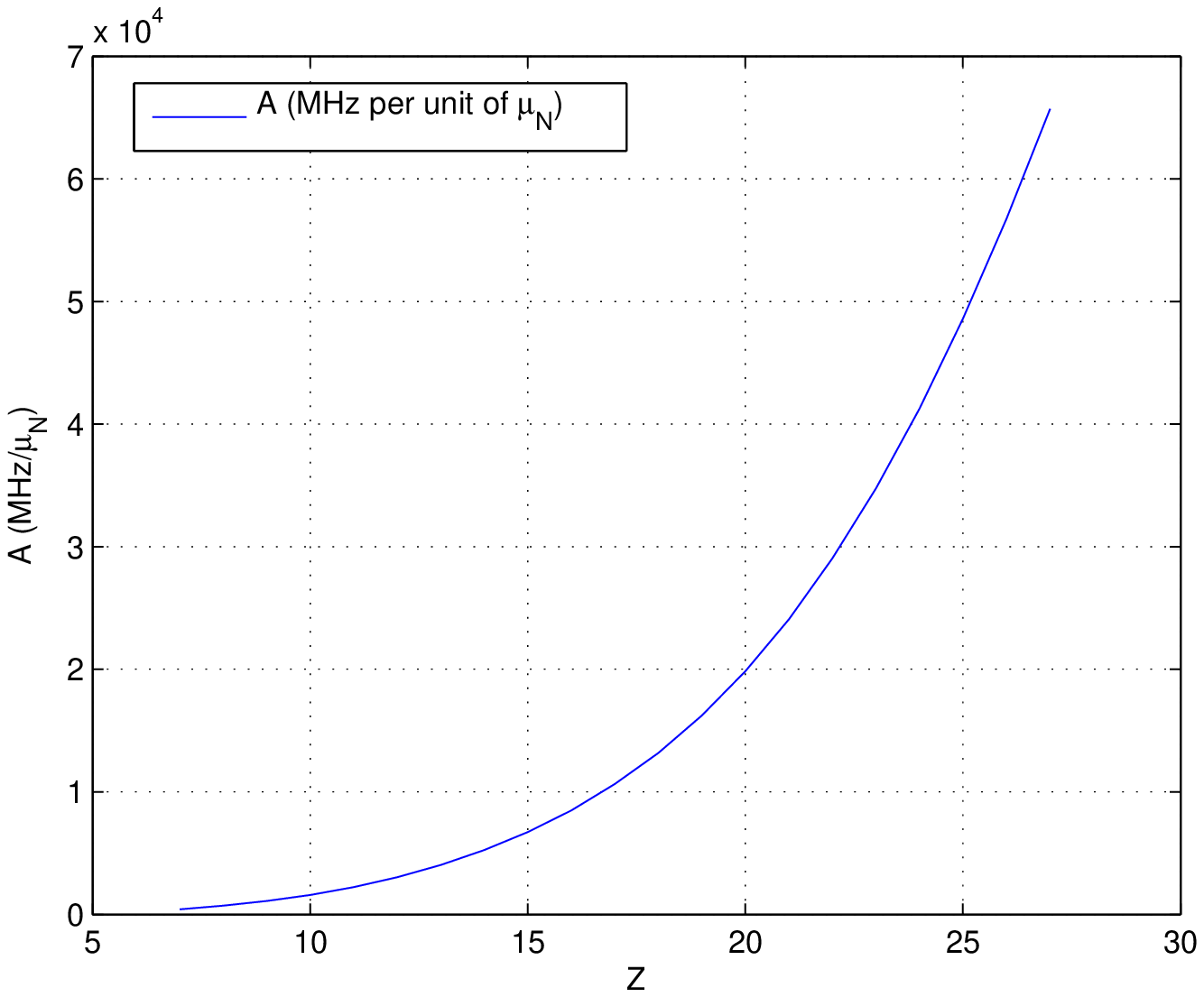}
\includegraphics[width=0.32\linewidth]{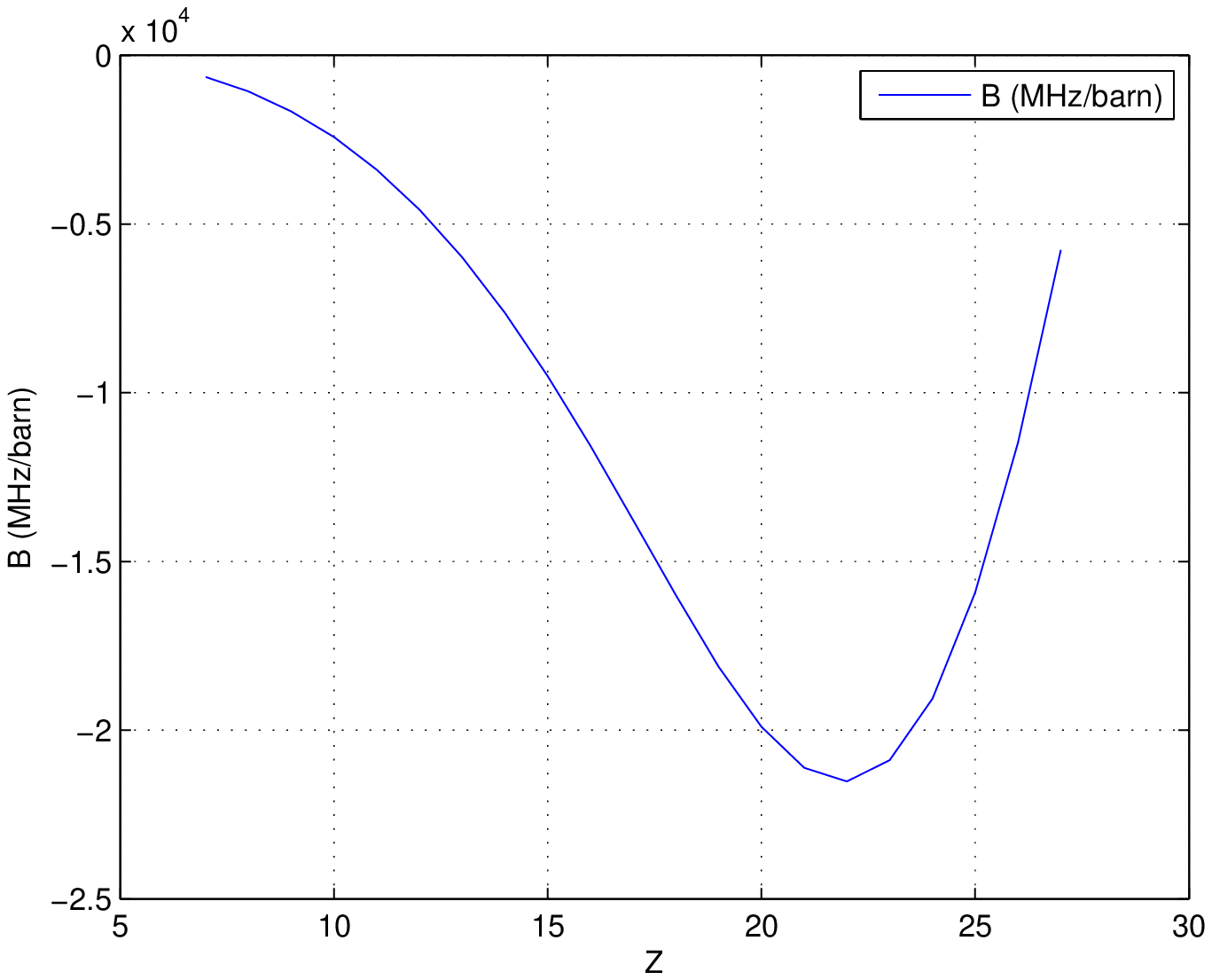}
%\end{minipage}
%\begin{minipage}[b!]{0.5\linewidth}
\includegraphics[width=0.32\linewidth]{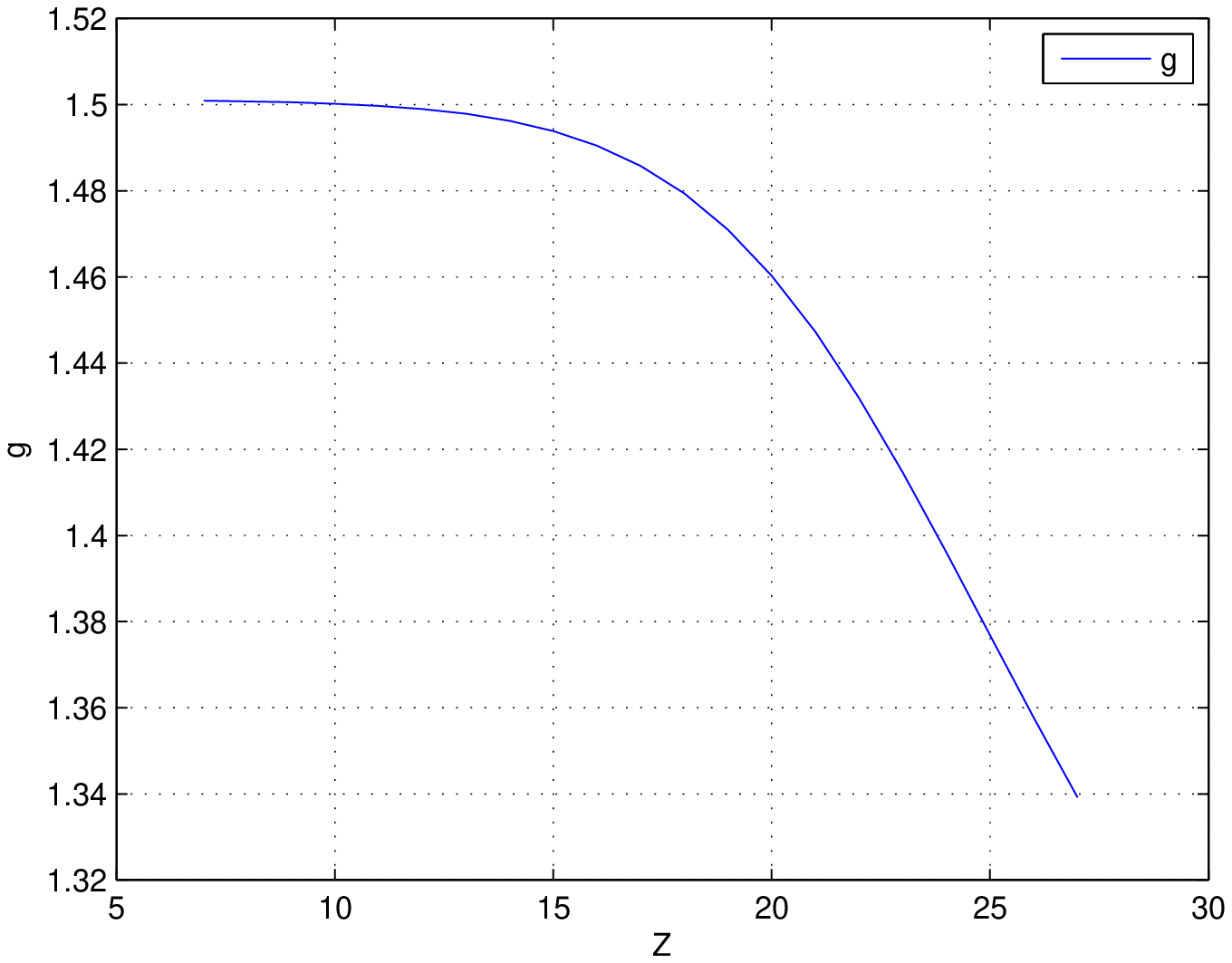}
\end{minipage}
\caption{ Evolution along the carbon isoelectronic sequence of the hyperfine constants $A$ (left), $B$ (middle), and the Land\'e $g_J$-factor (right)for the fourth root characterized by a $J=1$ total quantum number. This state presents a $1s^22s2p^3~^3S^o_{1}$ dominant character all along the sequence \label{fig:evolC-like}.}
\end{figure}

The reader should observe the slow drift, along the sequence, from a quite pure $LSJ$-coupling scheme (for low $Z$ value) to a $jj$ coupling scheme for the highly charged ions. This evolution along the isoelectronic sequence of the $1s^22s2p^3~^3S^o_{1}$ state is illustrated by 
figure~\ref{fig:evolC-like}. It is quite interesting to note the progressive decreasing of the $g_J$ factor when the nuclear charge increase. This behavior is explained by having a look at the second contributor in the $LSJ$ composition, i.e. $1s^22s2p^3~^1P^o_{1}$, which increases in importance for higher nuclear charges. 

J\"onsson and Biero\'n~\cite{JonBie:2010a} used the relativistic configuration interaction method to obtain energy levels, transition rates, hyperfine structure parameters and Land\'e $g_J$-factors for low-lying states in N~II, O~III, F~IV, Ne~V and Ti~XVII. Their values are not reported in this work, since they also used the \textsc{grasp2k} package to build them. The small relative differences compared with the present work, about 0.03\% and 0.3\% for, respectively, the $A_J$ and $B_J$ constants in Ti~XVII, come from differences in CSFs expansions and optimization strategies. To validate the hyperfine interaction constants from current RCI calculations we performed independent MCHF calculations for O III. 
Again, the MCHF calculations were fully variational, and the expansions were obtained by SD-excitations from large MR sets to an active set of orbitals with $nl \le9h$.
The MCHF results were relativistically corrected by scaling the interaction constants with the ratio of DHF and HF values.  The $A_J$ and $B_J$ constants are compared in table~\ref{tab:MCHF_Clike}. For all $A_J$ constants, except the one for the $1s^22s^22p^2~^3P_1$ state, there is a very good agreement between the RCI and MCHF calculations. The constants for $1s^22s^22p^2~^3P_1$ disagree with a factor of 2. This disagreement is due to a combination of severe cancellation effects, as discussed in section 4. In this case the $A_J$ constant from the MCHF calculation, that includes more correlation, is believed to be the more accurate. As $Z$ increases, electron correlation decrease in importance and the $A_J$ constant for the  $1s^22s^22p^2~^3P_1$ state from the RCI calculations gradually becomes more accurate. There are also very large differences between the $B_J$ constants based on RCI and MCHF for the states belonging to the $1s^22s2p^3$ configuration. Due to the half filled $2p$ subshell the $B_J$ constants are all zero at the HF level, making them sensitive to both polarization and higher order correlation effects as well as effects from relativistic mixing of states belonging to different $LS$ terms. 
Based on the fact that there is a good agreement between the two calculations for $1s^22s2p^3~^3D_3^o$, which is not affected by the $LS$ mixing, we draw the conclusion that 
the correlation effects have been reasonably captured in both the RCI and MCHF calculations. The differences for the other states are thus due to neglected relativistic $LS$ mixing in the MCHF calculations.
To conclude, $B_J$ constants for states belonging to the $1s^22s2p^3$ configuration are less accurate than the corresponding $A_J$ constants due to the fact that the HF values are zero, making correlation and relativistic effects comparatively more important. The $B_J$ values from the RCI calculations are the more accurate ones.

\subsection{Nitrogen-like systems}
Table~\ref{tab:HFS_N} presents the total and transition energies, the hyperfine interaction constants, the Land\'e $g_J$-factors, and the leading components of the many-electron wave function within the $LSJ$ coupling scheme. 
All figures of isoelectronic evolution for all states are available on demand from the corresponding author.

In Table~\ref{tab:MCHF_Nlike} the hyperfine interaction constants from the RCI calculations are compared with constants from large-scale fully variational MCHF calculations
in Ne IV. The MCHF results were relativistically corrected by scaling the interaction constants with the ratio of DHF and HF values. There is a very good agreement between 
the $A_J$ constants for all states except for $1s^22s^22p^3~^4S^o_{3/2}$. The $A_J$ constant for the latter state is zero at the HF level and, as discussed in~\cite{Godetal:97b}, higher order correlation effects are very important. The MCHF calculation includes more electron correlation and the corresponding  $A_J$ constant is believed to be more accurate than the constant from the RCI calculation. 
Turning to the $B_J$ constants we see that there is a perfect consistency for states belonging to the $1s^22s2p^4$ and $1s^22p^5$ configurations. However, for the two $J=3/2$ states belonging to $1s^22s^22p^3$, with 
a half filled $2p$ subshell, there is no agreement at all between the relativistic and non-relativistic calculations. Not even the signs are consistent. The fact that the non-relativistic MCHF calculations do not include effects of the $LS$ term mixing makes them virtually meaningless in this case. For the $J=5/2$ state there is no   $LS$ term mixing, and the $B_J$ constants from the RCI and MCHF calculations agree remarkably well, indicating that the correlation effects are indeed accurately described in both cases.

\ack
Part of this work was supported by the Communaut\'e fran\c caise of Belgium (Action de Recherche Concert\'ee), the Belgian National Fund for Scientific Research (FRFC/IISN Convention) and by the IUAP Belgian State Science Policy (BriX network P7/12).
CN and SV are grateful to the ``Fonds pour la formation ‡ la Recherche dans l'Industrie et dans l'Agriculture'' of Belgium for a PhD grant (Boursier F.R.S.-FNRS). PJ gratefully acknowledges financial support from the Swedish Research Council (VR). PJ and GG acknowledges support from the Visby program of the Swedish Institute.

\clearpage
%\bibliography{atoms_hfs}
%\bibliography{/Users/Ced/Documents/Dos_LaTeX/References/atoms}
%,atoms2}

\newpage

\TableExplanation
\bigskip
\renewcommand{\arraystretch}{1.0}

\section*{Table \ref{tab:HFS_Be}.\label{tbl1te} Total energies (in $E_{\text h}$), excitation energies (in cm$^{-1}$), hyperfine magnetic dipole constants $A_J(I/\mu_I)$ (MHz per unit of $\mu_N$), electric quadrupole constants $B_J/Q$ (MHz/barn) and Land\'e $g_J$-factors for levels in the beryllium isoelectronic sequence ($5\leq Z\leq 74$). For each of the many-electron wave functions, the leading components are given in the $LSJ$ coupling scheme. The number in square brackets is the power of 10.}
\begin{tabular*}{0.95\textwidth}{@{}@{\extracolsep{\fill}}lp{5.5in}@{}}
Level composition in $LSJ$ coupling & Leading components expressed within the $LSJ$ coupling scheme \\
								   & for each calculated level. \\
Energy          & The calculated absolute energies in atomic units ($E_{\text h}$).\\
$\Delta E$      & Calculated energies relatively to the lowest state reported in the table (in cm$^{-1}$).\\
$A_J(I/\mu_I)$& Magnetic dipole hyperfine interaction constant (MHz per units of $\mu_N$).\\
$B_J/Q$&Electric quadrupole hyperfine interaction constant (MHz/barn).\\
 $g_J$ &Land\'e $g_J$-factor.
\end{tabular*}
\label{tableI}

\section*{Table \ref{tab:Litzen}. Comparison of calculated hyperfine interaction constants $A_J(I/\mu_I)$ (in MHz per units of $\mu_N$) and $B_J/Q$ (MHz/barn) for the $2s 2p$ and $2p^2$ states in B II with Litz\'en \textit{et al.} \cite{Litetal:98a}.}
\begin{tabular*}{0.95\textwidth}{@{}@{\extracolsep{\fill}}lp{5.5in}@{}}
Config. &\\
$A_J(I/\mu_I)$& Magnetic dipole hyperfine interaction constants (MHz per units of $\mu_N$).\\
$B_J/Q$&Electric quadrupole hyperfine interaction constant (MHz/barn).\\
\end{tabular*}
\section*{Table \ref{tab:Marques}.\label{tbl1te} Comparison of the calculated hyperfine interaction constants $A_J(I/\mu_I)$ (MHz per unit of $\mu_N$) of the $ 1s^2 2s2p~^3P_1$ state for Be-like ions with the results obtained by Marques \emph{et al.}~\cite{Maretal:93a} and Cheng \emph{et al.}~\cite{Cheetal:2008a}.}
\begin{tabular*}{0.95\textwidth}{@{}@{\extracolsep{\fill}}lp{5.5in}@{}}
$A_J(I/\mu_I)$& Magnetic dipole hyperfine interaction constant (MHz per units of $\mu_N$).\\
$Z$ & nuclear charge.
\end{tabular*}
\section*{Table \ref{tab:HFS_B}.\label{tbl1te}  Total energies (in $E_{\text h}$), excitation energies (in cm$^{-1}$), hyperfine magnetic dipole constants $A_J(I/\mu_I)$ (MHz per unit of $\mu_N$), electric quadrupole constants $B_J/Q$ (MHz/barn) and Land\'e $g_J$-factors for levels in the boron isoelectronic sequence ($8\leq Z \leq  30$ and $Z=36,42$). For each of the many-electron wave functions, the leading components are given in the $LSJ$ coupling scheme. The number in square brackets is the power of 10.}
\begin{tabular*}{0.95\textwidth}{@{}@{\extracolsep{\fill}}lp{5.5in}@{}}
&Same as table  \ref{tab:HFS_Be}.
\end{tabular*}

\section*{Table \ref{tab:Dutta_Sun}. Comparison of calculated hyperfine interaction constants $A_J(I/\mu_I)$ (MHz per unit of $\mu_N$) and $B_J/Q$ (MHz/barn) with Dutta and Majumber \cite{DutMaj:2012a} and J\"onsson \textit{et al.} \cite{Jonetal:2010b}  for the O IV ion .}
\begin{tabular*}{0.95\textwidth}{@{}@{\extracolsep{\fill}}lp{5.5in}@{}}
$A_J(I/\mu_I)$& Magnetic dipole hyperfine interaction constant (MHz per units of $\mu_N$).\\
$B_J/Q$&Electric quadrupole hyperfine interaction constant (MHz/barn).\\
\end{tabular*}
\section*{Table \ref{tab:Dutta}. Comparison of calculated hyperfine interaction constants $A_J(I/\mu_I)$ (MHz per unit of $\mu_N$) with Dutta and Majumber \cite{DutMaj:2012a} for B-like ions from $Z=9$ to $Z=21$.}
\begin{tabular*}{0.95\textwidth}{@{}@{\extracolsep{\fill}}lp{5.5in}@{}}
$A_J(I/\mu_I)$& Magnetic dipole hyperfine interaction constant (MHz per units of $\mu_N$).\\
\end{tabular*}

\section*{Table \ref{tab:MCHF_OIV}. Hyperfine magnetic dipole constants $A_J(I/\mu_I)$ (MHz per unit of $\mu_N$) and electric quadrupole constants $B_J/Q$ (MHz/barn) in O~IV. Comparison of values from RCI and MCHF calculations.}
\begin{tabular*}{0.95\textwidth}{@{}@{\extracolsep{\fill}}lp{5.5in}@{}}
&Same as table~\ref{tab:Dutta_Sun}.\\
\end{tabular*}

\section*{Table \ref{tab:HFS_C}.\label{tbl1C} Total energies (in $E_{\text h}$), excitation energies (in cm$^{-1}$), hyperfine magnetic dipole constants $A_J(I/\mu_I)$ (MHz per unit of $\mu_N$), electric quadrupole constants $B_J/Q$ (MHz/barn) and Land\'e $g_J$-factors for levels in the carbon isoelectronic sequence ($7\leq Z \leq28$). For each of the many-electron wave functions, the leading components are given in the $LSJ$ coupling scheme. The number in square brackets is the power of 10.}
\begin{tabular*}{0.95\textwidth}{@{}@{\extracolsep{\fill}}lp{5.5in}@{}}
&Same as table  \ref{tab:HFS_Be}.
\end{tabular*}
\section*{Table \ref{tab:MCHF_Clike}. Hyperfine magnetic dipole constants $A_J(I/\mu_I)$ (MHz per unit of $\mu_N$) and electric quadrupole constants $B_J/Q$ (MHz/barn) in O~III. Comparison of values from RCI and MCHF calculations.}
\begin{tabular*}{0.95\textwidth}{@{}@{\extracolsep{\fill}}lp{5.5in}@{}}
&Same as table~\ref{tab:Dutta_Sun}.\\
\end{tabular*}
\section*{Table \ref{tab:HFS_N}. 
Total energies (in $E_{\text h}$), excitation energies (in cm$^{-1}$), hyperfine magnetic dipole constants $A_J(I/\mu_I)$ (MHz per unit of $\mu_N$), electric quadrupole constants $B_J/Q$ (MHz/barn) and Land\'e $g_J$-factors for levels in the nitrogen isoelectronic sequence ($7\leq Z\leq36$ and $Z=42,74$). For each of the many-electron wave functions, the leading components are given in the $LSJ$ coupling scheme. The number in square brackets is the power of 10.\label{tbl1N} }
\begin{tabular*}{0.95\textwidth}{@{}@{\extracolsep{\fill}}lp{5.5in}@{}}
&Same as table  \ref{tab:HFS_Be}.
\end{tabular*}
\section*{Table \ref{tab:MCHF_Nlike}. Hyperfine magnetic dipole constants $A_J(I/\mu_I)$ (MHz per unit of $\mu_N$) and electric quadrupole constants $B_J/Q$ (MHz/barn) in Ne~IV. Comparison of values from RCI and MCHF calculations.}
\begin{tabular*}{0.95\textwidth}{@{}@{\extracolsep{\fill}}lp{5.5in}@{}}
&Same as table~\ref{tab:Dutta_Sun}.\\
\end{tabular*}
%\section*{Table \ref{tab:God}.Comparison with other works for the hyperfine structure coupling constant of the level $2s^22p^3~^4S^o_{3/2}$ in $^{14}$N.}
%\begin{tabular*}{0.95\textwidth}{@{}@{\extracolsep{\fill}}lp{5.5in}@{}}
%Config. &\\
%\end{tabular*}
\datatables % This command is necessary to get the table names in toc

%% One-page data tables are also best formatted using the longtable
%% environment:
%\begin{longtable}{c}
%\caption{This is the First Data Table}\\
%\endhead\\
%\end{longtable}

%% If the table is to span over the whole text width, we set:

\setlength{\LTleft}{0pt}
\setlength{\LTright}{0pt} 

% To avoid ``Overfull \hboxes...'' decrease the intercolumn spacing:

\setlength{\tabcolsep}{0.5\tabcolsep}

\renewcommand{\arraystretch}{1.0}

\footnotesize % we need to squeeze the font size a lot!

\newpage

\begin{longtable}{@{\extracolsep\fill}llllllrrrc@{}}
\caption{\label{tab:HFS_Be} 
Total energies (in $E_{\text h}$), excitation energies (in cm$^{-1}$), hyperfine magnetic dipole constants $A_J(I/\mu_I)$ (MHz per unit of $\mu_N$), electric quadrupole constants $B_J/Q$ (MHz/barn), and Land\'e $g_J$-factors for levels in the beryllium isoelectronic sequence ($5\leq Z\leq 74$). For each of the many-electron wave functions, the leading components are given in the $LSJ$ coupling scheme. The number in square brackets is the power of 10. See page \pageref{tbl1te} for Explanation of Tables.}\\
\multicolumn{4}{l}{Level composition in $LSJ$ coupling}&$J$& Energy ($E_{\rm h}$) &$\Delta E$ cm$^{-1}$& \multicolumn{1}{c}{$A_J(I/\mu_I)$} & \multicolumn{1}{c}{$B_J/Q$}& $g_J$\\
\endfirsthead\\
\caption[]{(continued)}\\ [-.4cm]
\multicolumn{4}{l}{Level composition in $LSJ$ coupling}&$J$& Energy ($E_{\rm h}$) &$\Delta E$ cm$^{-1}$& \multicolumn{1}{c}{$A_J(I/\mu_I)$} & \multicolumn{1}{c}{$B_J/Q$}& $g_J$\\ 
\hline
\endhead
\hline&&&& \multicolumn{3}{c}{B~II} \\ \hline
$1s^22s2p$ $^3P^o$ &(99\%) &&&1 &$ -24.1835957 $&$37277.92 $&$5.132[2]$& $-5.247[1]$& 1.501085 \\ 
&(99\%) &&&2 &$ -24.1835226 $&$37293.97 $&$4.428[2]$& $1.049[2]$& 1.501095 \\ 
$1s^22s2p$ $^1P^o$ &(96\%) &+ $1s^22p3d$ $^1P^o$ &(3\%) &1 &$ -24.0185813 $&$73494.38 $&$1.015[2]$& $9.441[1]$& 0.999946 \\ 
$1s^22p^2$ $^3P$ &(99\%) &&&1 &$ -23.9028320 $&$98898.43 $&$-6.309[1]$& $5.183[1]$& 1.501107 \\ 
&(99\%) &&&2 &$ -23.9027712 $&$98911.77 $&$6.098 $& $-1.038[2]$& 1.501091 \\ 
$1s^22p^2$ $^1D$ &(93\%) &+ $1s^22s3d$ $^1D$ &(6\%) &2 &$ -23.8868692 $&$102401.85 $&$1.001[2]$& $1.887[2]$& 0.999954 \\ 
\hline&&&& \multicolumn{3}{c}{C~III} \\
\end{longtable}

\newpage

\begin{table}%[htdp]
\caption{\label{tab:Litzen}Be-like boron: comparison of  hyperfine interaction constants $A_J(I/\mu_I)$ (MHz per units of $\mu_N$) and $B_J/Q$ (MHz/barn) with Litz\'en \textit{et al.}~\cite{Litetal:98a}  for the $2s 2p$ and $2p^2$ states.} 
\begin{tabular}{@{}l|rr|rr|rr|rr|} 
Designation&\multicolumn{2}{c}{$A_1(I/\mu_I)$}&\multicolumn{2}{c}{$A_2(I/\mu_I)$}&\multicolumn{2}{c}{$B_1/Q$}&\multicolumn{2}{c}{$B_2/Q$}\\\\
&Litz\'en \textit{et al.}&This work&Litz\'en \textit{et al.}&This work&Litz\'en \textit{et al.}&This work&Litz\'en \textit{et al.}&This work\\
\hline
$2s2p~^3P^o$&512.60	&513.2&443.00&442.8&$-51.32$&$-52.47$&102.7&104.9 \\
$2s2p~^1P^o$&101.20	&101.5&	&   		&	$94.23$&94.41&&\\
$2p^2~^3P$&$-62.819$	&$-63.09$&6.081&6.098&51.13 &51.83&$- 102.3$&$-103.8$\\
$2p^2~^1D$&			& 	&99.859&100.1&&&189.0 &188.7\\
\hline\hline
\end{tabular}
\end{table}

\newpage

\begin{table}
\centering
\caption{\label{tab:Marques}Comparison of calculated hyperfine interaction constants $A_J(I/\mu_I)$ (MHz per unit of $\mu_N$) of the $ 1s^2 2s2p~^3P^o_1$ state for Be-like ions with the results obtained by Marques \emph{et al.}~ \cite{Maretal:93a} and Cheng \emph{et al.}~\cite{Cheetal:2008a}.}
\begin{tabular}{lc|ccc||c|ccccc}
&\multicolumn{8}{c}{$A_J(I/\mu_I)$}\\
\hline
Z&&Marques \textit{et  al.} \cite{Maretal:93a}&Cheng \textit{et  al.} \cite{Cheetal:2008a}&This work
&Z&&Marques \textit{et  al.} \cite{Maretal:93a}&Cheng \textit{et  al.} \cite{Cheetal:2008a}&This work\\
%\hline
%\endhead
\hline
5      & &4.669[2]   &     &5.132[2] 		&40    & &1.347[6]&1.3599[6] &1.365[6]\\
6	& &$1.040[3]$&1.1151[3]&1.114[3]	&41    & &1.479[6]&1.4935[6] &1.499[6]\\
7	& &$1.948[3]$&2.0558[3]&2.056[3]	&42    & &1.621[6]&1.6362[6] &1.642[6]\\
8     & &$3.269[3]$&3.4158[3]&3.417[3]	&43    & &1.773[6]&1.7884[6] &1.795[6]\\
9	& &$5.082[3]$&5.2782[3]&5.280[3]	&44    & &1.934[6]&1.9504[6] &1.958[6]\\
10    & & 7.481[3]&7.7328[3]&7.736[3]	&45	& &$2.106[6]$&2.1230[6] &2.132[6] \\
\hline\hline\end{tabular}
\end{table}

\clearpage

\begin{longtable}{@{\extracolsep\fill}llllllrrrc@{}}
\caption{\label{tab:HFS_B} 
Total energies (in $E_{\text h}$), excitation energies (in cm$^{-1}$), hyperfine magnetic dipole constants $A_J(I/\mu_I)$ (MHz per unit of $\mu_N$), electric quadrupole constants $B_J/Q$ (MHz/barn), and Land\'e $g_J$-factors for levels in the boron isoelectronic sequence ($8\leq Z \leq  30$ and $Z=36,42$). For each of the many-electron wave functions, the leading components are given in the $LSJ$ coupling scheme. The number in square brackets is the power of 10. See page \pageref{tbl1te} for Explanation of Tables. }\\
\multicolumn{4}{l}{Level composition in $LSJ$ coupling}&$J$& Energy ($E_{\rm h}$) &$\Delta E$ cm$^{-1}$& \multicolumn{1}{c}{$A_J(I/\mu_I)$} & \multicolumn{1}{c}{$B_J/Q$}& $g_J$\\
\endfirsthead\\
\caption[]{(continued)}\\ [-.4cm]
\multicolumn{4}{l}{Level composition in $LSJ$ coupling}&$J$& Energy ($E_{\rm h}$) &$\Delta E$ cm$^{-1}$& \multicolumn{1}{c}{$A_J(I/\mu_I)$} & \multicolumn{1}{c}{$B_J/Q$}& $g_J$\\ 
\hline
\endhead
\hline&&&& \multicolumn{3}{c}{O~IV} \\ \hline
$1s^22s^22p$ $^2P^o$ &(95\%) &+ $1s^22p^3$ $^2P^o$ &(4\%) &1/2 &$ -71.3006468 $&$0.00 $&$2.174[3]$& $0.0$& 0.665627 \\ 
&(95\%) &+ $1s^22p^3$ $^2P^o$ &(4\%) &3/2 &$ -71.2988722 $&$389.49 $&$4.262[2]$& $7.426[2]$& 1.333886 \\ 
$1s^22s2p^2$ $^4P$ &(100\%) &&&1/2 &$ -70.9755351 $&$71353.76 $&$4.474[3]$& $0.0$& 2.670161 \\ 
&(100\%) &&&3/2 &$ -70.9749349 $&$71485.49 $&$2.083[3]$& $6.101[2]$& 1.734786 \\ 
&(100\%) &&&5/2 &$ -70.9740914 $&$71670.63 $&$2.115[3]$& $-7.643[2]$& 1.601127 \\ 
$1s^22s2p^2$ $^2D$ &(98\%) &+ $1s^22s^23d$ $^2D$ &(1\%) &5/2 &$ -70.7218546 $&$127030.20 $&$2.432[3]$& $1.490[3]$& 1.200243 \\ 
&(98\%) &+ $1s^22s^23d$ $^2D$ &(1\%) &3/2 &$ -70.7217904 $&$127044.30 $&$-8.796[2]$& $1.043[3]$& 0.799362 \\
\end{longtable}

\newpage

\begin{table}%[htdp]
\centering
\caption{\label{tab:Dutta_Sun}Comparison of calculated hyperfine interaction constants $A_J(I/\mu_I)$ (MHz per unit of $\mu_N$) and $B_J/Q$ (MHz/barn) with Dutta and Majumber \cite{DutMaj:2012a}, J\"onsson \textit{et al.} \cite{Jonetal:2010b}, and Sun \textit{et al.} \cite{Sunetal:2012a}  for the O IV ion. RCI are results from this work.} 
\begin{tabular}{@{}r|rrr|rrrrrrrrrr} 
\multicolumn{1}{l|}{Level}&\multicolumn{3}{c|}{$A_J(I/\mu_I)$}&\multicolumn{4}{c}{$B_J/Q$}\\
%&&&&&\\
& Ref. \cite{DutMaj:2012a}&Ref. \cite{Jonetal:2010b}&RCI\hspace{1cm}&Ref. \cite{Jonetal:2010b}&&RCI&&&\\
\hline
$1s^22s^22p~^2P^o_{1/2}$	&2192&2174&2174&0.0&&0.0 &&&&\\
			$^2P^o_{3/2}$	&453&427.7&426.2&745&&742.6 &&&&\\
\hline&&&&\\
&Ref. \cite{Sunetal:2012a}&Ref. \cite{Jonetal:2010b}&RCI&Ref. \cite{Sunetal:2012a}&Ref. \cite{Jonetal:2010b}&RCI\\ 
\hline
$1s^22s2p^2$ $^4P_{5/2}$ &$2 157.6$&2116&$\phantom{-}2115$	&$-760$&					&$-764.27$	\\ 
			$^4P_{3/2}$&$2 135.9$&2083&$\phantom{-}2083$	&$609$&$\phantom{-}610.1$	&611.40		\\ 
			$^4P_{1/2}$&$4 592.4$&4475&$\phantom{-}4474$	&0.0&$\phantom{-}0.0$			&0.0			\\ 
\hline
\end{tabular}
\end{table}

\newpage

\begin{table}%[htdp]
\centering
\caption{\label{tab:Dutta}Comparison of calculated hyperfine interaction constants $A_J(I/\mu_I)$ (MHz per unit of $\mu_N$) and $B_J/Q$ (MHz/barn) with Dutta and Majumber \cite{DutMaj:2012a} for B-like ions from $Z=9$ to $Z=21$. The number in square brackets is the power of 10. } 
\begin{tabular}{@{}lr|cr@{}} 
\hline
&&\multicolumn{2}{c}{$A_J(I/\mu_I)$}\\
\hline%\cline{3-4}
Z&Level&\multicolumn{1}{c}{Dutta and Majumber \cite{DutMaj:2012a}}&This work\\
\hline
%8	&$2s^22p~^2P^o_{1/2}$&2192	&2.174[+3]\\
%	&$2s^22p~^2P^o_{3/2}$&452		&4.26.6[+2]\\
9	&$2p~^2P^o_{1/2}$	&\phantom{9}3.5569[3]	&3.531[3]\\
	&$ ~^2P^o_{3/2}$	&\phantom{469}7.30[2]	&6.902[2]\\
10	&$2p~^2P^o_{1/2}$	&\phantom{69}5.383[3]	&5.348[3]\\
	&$ ~^2P^o_{3/2}$	&\phantom{469}1.10[3]	&1.043[3]\\
11	&$2p~^2P^o_{1/2}$	&\phantom{6}7.7381[3]	&7.692[3]\\
	&$ ~^2P^o_{3/2}$	&\phantom{69}1.576[3]	&1.498[3]\\
12	&$2p~^2P^o_{1/2}$	&\phantom{6}1.0696[4]	&1.064[4]\\
	&$ ~^2P^o_{3/2}$	&\phantom{469}2.17[3]	&2.069[3]\\
13	&$2p~^2P^o_{1/2}$	&\phantom{6}1.4321[4]	&1.424[4]\\
	&$ ~^2P^o_{3/2}$	&\phantom{69}2.903[3]	&2.766[3]\\
14	&$2p~^2P^o_{1/2}$	&\phantom{6}1.8689[4]	&1.859[4]\\
	&$ ~^2P^o_{3/2}$	&\phantom{69}3.779[3]	&3.604[3]\\
15	&$2p~^2P^o_{1/2}$	&\phantom{6}2.3873[4]	&2.375[4]\\
	&$ ~^2P^o_{3/2}$	&\phantom{6}4.8149[3]	&4.596[3]\\
16	&$2p~^2P^o_{1/2}$	&\phantom{6}2.9948[4]	&2.979[4]\\
	&$ ~^2P^o_{3/2}$	&\phantom{69}6.025[3]	&5.755[3]\\
17	&$2p~^2P^o_{1/2}$	&\phantom{6}3.6988[4]	&3.680[4]\\
	&$~^2P^o_{3/2}$	&\phantom{69}7.419[3]	&7.093[3]\\
18	&$2p~^2P^o_{1/2}$	&\phantom{6}4.5078[4]	&4.485[4]\\
	&$~^2P^o_{3/2}$	&\phantom{69}9.015[3]	&8.624[3]\\
19	&$2p~^2P^o_{1/2}$	&\phantom{6}5.4292[4]	&5.403[4]\\
	&$~^2P^o_{3/2}$	&\phantom{6}1.0823[4]	&1.036[4]\\
20	&$2p~^2P^o_{1/2}$	&\phantom{6}6.4722[4]	&6.442[4]\\
	&$~^2P^o_{3/2}$	&\phantom{6}1.2855[4]	&1.232[4]\\
21	&$2p~^2P^o_{1/2}$	&7.64496[4]	&7.610[4]\\
	&$~^2P^o_{3/2}$	&\phantom{6}1.5127[4]	&1.451[4]\\
\hline
\end{tabular}
\end{table}

\newpage

\begin{table}\centering
\caption{\label{tab:MCHF_OIV}Hyperfine magnetic dipole constants $A_J(I/\mu_I)$ (MHz per unit of $\mu_N$) and electric quadrupole constants $B_J/Q$ (MHz/barn) in O IV. Comparison of values from RCI and MCHF calculations.}
\begin{tabular}{r|rr|rrrr} \hline
                       & \multicolumn{2}{c}{RCI (this work)} && \multicolumn{2}{c}{MCHF (this work)} \\ 
                       \cline{2-3}\cline{5-6}
Level                 &\multicolumn{1}{c}{$A_J(I/\mu_I)$}  &\multicolumn{1}{c}{$B_J/Q$}&&
\multicolumn{1}{c}{$A_J(I/\mu_I)$}&\multicolumn{1}{c}{$B_J/Q$}\\
\hline
$2s^22p~^2P^o_{1/2}$   	&  2174\phantom{.0}&		&& 2175\phantom{.0}&           \\      
     $~^2P^o_{3/2}$   	&  426.2    & 742.6	&& 427.9    &    743.5  \\&&&&\\
$2s2p^2~^4P_{1/2}$    	&  4474\phantom{.0}&		&& 4473\phantom{.0}&           \\
      $~^4P_{3/2}$    		&  2083\phantom{.0}& 610.1	&& 2081\phantom{.0}&    610.7  \\
      $~^4P_{5/2}$    		&  2115\phantom{.0}&$-764.3$&& 2115\phantom{.0}&   $-765.0$ \\&&&&\\
$2s2p^2~^2D_{3/2}$    	&$-879.6$ & 1043\phantom{.0}&& $-879.4$ &   1042\phantom{.0}\\
      $~^2D_{5/2}$    		&  2432\phantom{.0}& 1490\phantom{.0}&& 2431\phantom{.0}& 1489\phantom{.0}\\&&&&\\
$2s2p^2~^2P_{1/2}$    	&  1438\phantom{.0}&             && 1437\phantom{.0}&           \\
      $~^2P_{3/2}$    		&$-514.6$ &$-765.7$&& $-513.1$ &  $-765.6$ \\&&&&\\
$2s2p^2~^2S_{1/2}$    	&  9060\phantom{.0}&		&& 9051\phantom{.0}&\\&&&&\\ 
$2p^3~^4S^o_{3/2}$    	&$-466.7$ &$-0.0002$&& $-466.5$ &  0\phantom{.0}\\&&&&\\
$2p^3~^2D^o_{3/2}$    	&  548.6    &$-15.3$	&& 547.7    &       $-4.5$ \\
    $~^2D^o_{5/2}$    	&  648.4    &$-43.9$	&& 648.30   &       $-44.4$ \\&&&&\\
$2p^3~^2P^o_{1/2}$    	&  2232\phantom{.0}&		&& 2230.8   &               \\
    $~^2P^o_{3/2}$    	&  258.3    &44.6	&& 258.9    &    35.0 \\ \hline
\end{tabular}
\end{table}
\clearpage

\begin{longtable}{@{\extracolsep\fill}llllllrrrc@{}}
\caption{\label{tab:HFS_C}Total energies (in $E_{\text h}$), excitation energies (in cm$^{-1}$), hyperfine magnetic dipole constants $A_J(I/\mu_I)$ (MHz per unit of $\mu_N$), electric quadrupole constants $B_J/Q$ (MHz/barn), and Land\'e $g_J$-factors for levels in the carbon isoelectronic sequence ($7\leq Z \leq28$). For each of the many-electron wave functions, the leading components are given in the $LSJ$ coupling scheme. The number in square brackets is the power of 10. See page \pageref{tbl1C} for Explanation of Tables.}\\
\multicolumn{4}{l}{Level composition in $LSJ$ coupling}&$J$& Energy ($E_{\rm h}$) &$\Delta E$ cm$^{-1}$& \multicolumn{1}{c}{$A_J(I/\mu_I)$} & \multicolumn{1}{c}{$B_J/Q$}& $g_J$\\
\endfirsthead\\
\caption[]{(continued)}\\ [-.4cm]
\multicolumn{4}{l}{Level composition in $LSJ$ coupling}&$J$& Energy ($E_{\rm h}$) &$\Delta E$ cm$^{-1}$& \multicolumn{1}{c}{$A_J(I/\mu_I)$} & \multicolumn{1}{c}{$B_J/Q$}& $g_J$\\ 
\hline
\endhead
\hline&&&& \multicolumn{3}{c}{N~II} \\ \hline
$1s^22s^22p^2$ $^3P$ &(96\%) &+ $1s^22p^4$ $^3P$ &(2\%) &1 &$ -54.0761997 $&$50.41 $&$-3.140 $& $1.693[2]$& 1.501039 \\ 
&(96\%) &+ $1s^22p^4$ $^3P$ &(2\%) &2 &$ -54.0758321 $&$131.07 $&$2.297[2]$& $-3.394[2]$& 1.501007 \\ 
$1s^22s^22p^2$ $^1D$ &(96\%) &+ $1s^22p^4$ $^1D$ &(2\%) &2 &$ -54.0059995 $&$15457.56 $&$3.578[2]$& $6.596[2]$& 0.999896 \\ 
$1s^22s2p^3$ $^5S^o$ &(99\%) &&&2 &$ -53.8639833 $&$46626.52 $&$1.122[3]$& $-6.192[-3]$& 2.002150 \\ 
$1s^22s2p^3$ $^3D^o$ &(96\%) &+ $1s^22s^22p3d$ $^3D^o$ &(2\%) &3 &$ -53.6555256 $&$92377.69 $&$1.061[3]$& $-3.899[1]$& 1.333967 \\ 
&(96\%) &+ $1s^22s^22p3d$ $^3D^o$ &(2\%) &2 &$ -53.6554696 $&$92389.98 $&$5.924[2]$& $-1.421[1]$& 1.166948 \\ 
&(96\%) &+ $1s^22s^22p3d$ $^3D^o$ &(2\%) &1 &$ -53.6554634 $&$92391.35 $&$-9.073[2]$& $-1.100[1]$& 0.498815 \\
\end{longtable}

\newpage

\begin{table}\centering
\caption{\label{tab:MCHF_Clike}Hyperfine magnetic dipole constants $A_J(I/\mu_I)$ (MHz per unit of $\mu_N$) and electric quadrupole constants $B_J/Q$ (MHz/barn) in O III. Comparison of values from RCI and MCHF calculations.}
\begin{tabular}{r|rrrrrr} \hline
                       & \multicolumn{2}{c}{RCI (this work)} && \multicolumn{2}{c}{MCHF (this work)} \\ 
                       \cline{2-3}\cline{5-6}
Level                 &\multicolumn{1}{c}{$A_J(I/\mu_I)$}  &\multicolumn{1}{c}{$B_J/Q$}&&
\multicolumn{1}{c}{$A_J(I/\mu_I)$}&\multicolumn{1}{c}{$B_J/Q$}\\
\hline
$1s^22s^22p^2$ $^3P_1$	& $-7.847 $& $3.188[2]$ 		&& $-3.426$ & $3.193[2]$\\
$^3P_2$					&$4.251[2]$& $-6.392[2]$ 	&&4.242[2] &$-6.387[2]$\\\\
$1s^22s^22p^2$ $^1D_2$	&$6.709[2]$& $1.249[3]$ 		&\phantom{6.694[2]}& $6.694[2]$ & $1.270[3]$\\\\
$1s^22s2p^3$ $^5S^o_2$ 	&$1.906[3]$& $-1.052[-2]$ 	&& $1.897[3]$ & $0.0$ \\\\
$1s^22s2p^3$ $^3D^o_3$	&$1.856[3]$& $-4.271[1]$ 	&& $1.851[3]$ &$ -4.396[1]$\\
$^3D^o_2$				&$1.047[3]$& $-3.447 $ 		&& $1.045[3]$ & $-2.198[1]$\\
$^3D^o_1$				&$-1.487[3]$& $-6.019 $ 		&&$-1.483[3]$ & $-1.539[1]$\\
\\$1s^22s2p^3$ $^3P^o_2$	&$2.240[3]$& $1.720[1]$ 		&& $2.227[3]$ & $3.422[1]$\\
$^3P^o_1$				&$2.660[3]$& $-2.673[1]$ 	&& $2.650[3]$ &$-1.711[1]$\\
\\$1s^22s2p^3$ $^1D^o_2$	&$6.577[2]$& $-1.635[1]$ 	&& $6.547[2]$ &$-1.854[1]$\\
\\$1s^22s2p^3$ $^3S^o_1$	&$-2.304[3]$& $-8.427[-3]$ 	&& $-2.291[3]$ & $0.0$ \\
\\$1s^22s2p^3$ $^1P^o_1$	&$6.739[2]$& $2.581[1]$ 		&& $6.763[2]$ & $3.201[1]$\\
\\$1s^22p^4$ $^3P_2$		&$2.102[2]$& $5.451[2]$ 		&& $2.228[2]$  & $5.378[2]$\\
$^3P_1$					&$-1.854[2]$& $-2.721[2]$ 	&&$-1.726[2]$&$-2.689[2]$\\
\\$1s^22p^4$ $^1D_2$		&$6.513[2]$& $-1.178[3]$ 	&& $6.478[2]$ & $-1.205[3]$\\
\end{tabular}
\end{table}
\footnotesize

\clearpage
\begin{longtable}{@{\extracolsep\fill}llllllrrrc@{}}
\caption{\label{tab:HFS_N}
Total energies (in $E_{\text h}$), excitation energies (in cm$^{-1}$), hyperfine magnetic dipole constants $A_J(I/\mu_I)$ (MHz per unit of $\mu_N$), electric quadrupole constants $B_J/Q$ (MHz/barn), and Land\'e $g_J$-factors for levels in the nitrogen isoelectronic sequence ($7\leq Z\leq36$ and $Z=42,74$). For each of the many-electron wave functions, the leading components are given in the $LSJ$ coupling scheme. The number in square brackets is the power of 10. See page \pageref{tbl1N} for Explanation of Tables.}\\
\multicolumn{4}{l}{Level composition in $LSJ$ coupling}&$J$& Energy ($E_{\rm h}$) &$\Delta E$ cm$^{-1}$& \multicolumn{1}{c}{$A_J(I/\mu_I)$} & \multicolumn{1}{c}{$B_J/Q$}& $g_J$\\
\endfirsthead\\
\caption[]{(continued)}\\ [-.4cm]
\multicolumn{4}{l}{Level composition in $LSJ$ coupling}&$J$& Energy ($E_{\rm h}$) &$\Delta E$ cm$^{-1}$& \multicolumn{1}{c}{$A_J(I/\mu_I)$} & \multicolumn{1}{c}{$B_J/Q$}& $g_J$\\ 
\hline
\endhead
\hline&&&& \multicolumn{3}{c}{N~I} \\ \hline
$2s^22p^3$ $^4S^o$ &(96\%) &+ $1s^22s2p^33d$ $^4S^o$ &(2\%) &3/2 &$ -54.6073794 $&$0.00 $&$1.385[1]$& $-1.496[-3]$& 1.999857 \\ 
$2s^22p^3$ $^2D^o$ &(96\%) &+ $1s^22s2p^33d$ $^2D^o$ &(1\%) &5/2 &$ -54.5185543 $&$19494.84 $&$2.825[2]$& $-2.134[1]$& 1.199893 \\ 
&(96\%) &+ $1s^22s2p^33d$ $^2D^o$ &(1\%) &3/2 &$ -54.5185222 $&$19501.91 $&$1.670[2]$& $-9.281 $& 0.799970 \\ 
$2s^22p^3$ $^2P^o$ &(93\%) &+ $2p^5$ $^2P^o$ &(3\%) &3/2 &$ -54.4735477 $&$29372.66 $&$1.534[2]$& $5.870 $& 1.333197 \\ 
&(93\%) &+ $2p^5$ $^2P^o$ &(3\%) &1/2 &$ -54.4735265 $&$29377.32 $&$7.955[2]$& $0.0$& 0.666550 \\ 
$2s2p^4$ $^4P$ &(91\%) &+ $1s^22s^22p^23d$ $^4P$ &(3\%) &5/2 &$ -54.2058086 $&$88134.61 $&$9.117[2]$& $2.156[2]$& 1.601263 \\ 
&(91\%) &+ $1s^22s^22p^23d$ $^4P$ &(3\%) &3/2 &$ -54.2055167 $&$88198.66 $&$9.491[2]$& $-1.721[2]$& 1.734907 \\
\end{longtable}

%\newpage

%\begin{table}%[htdp]
%\caption{\label{tab:God}Comparison with other works for the hyperfine structure coupling constant of the level$2s^22p^3~^4S^o_{3/2}$ in $^{14}$N.} 
%\begin{tabular}{@{}rr|rrr} 
%Reference&&\multicolumn{2}{c}{$A_{3/2}$ (MHz)}&$B_{3/2}$ (Hz)\\
%%&\multicolumn{3}{c}{$B_J$}\\
%\hline
%Godefroid et al. \cite{Godetal:97b}&SDTQ-CI	&10.40&&\\
%&SD-MCHF&5.66\\
%This work&&5.59&&29.93\\
%&\\
%Hirsch \textit{et al.} \cite{Hiretal:77a}	&Exp.	&10.4509&&1.27(10)\\
%\hline
%\\
%\hline\hline
%\end{tabular}
%\end{table}
\newpage

\begin{table}\centering
\caption{\label{tab:MCHF_Nlike}Hyperfine magnetic dipole constants $A_J(I/\mu_I)$ (MHz per unit of $\mu_N$) and electric quadrupole constants $B_J/Q$ (MHz/barn) in Ne~IV. Comparison of values from RCI and MCHF calculations.}
\begin{tabular}{r|rrrrrr} \hline
                       & \multicolumn{2}{c}{RCI (this work )} && \multicolumn{2}{c}{MCHF (this work )} \\ 
                       \cline{2-3}\cline{5-6}
Level                 &\multicolumn{1}{c}{$A_J(I/\mu_I)$}  &\multicolumn{1}{c}{$B_J/Q$}&&
\multicolumn{1}{c}{$A_J(I/\mu_I)$}&\multicolumn{1}{c}{$B_J/Q$}\\
\hline
$2s^22p^3$ $^4S^o_{3/2}$	&$2.735[1]$&	$-3.420[-2]$	&\phantom{1.501}& 2.093[1]		& 0.0	\\\\
$2s^22p^3$ $^2D^o_{5/2}$	&$1.501[3]$&	$-4.492[1]$	&& 1.501[3] 	& $-4.486[1]$ \\
$^2D^o_{3/2}$				&$9.349[2]$&	$1.083[2]$	&&	9.157[2] 	&$-3.141[1]$	\\\\
$2s^22p^3$ $^2P^o_{1/2}$ 	&$4.222[3]$&	$0.0$		&&	4.187[3] 	&	0.0\\
$^2P^o_{3/2}$				&$8.174[2]$&	$-1.287[2]$	&& 8.195[2]	& 1.458[1]\\\\
$2s2p^4$ $^4P_{5/2}$ 		&$4.041[3]$&	$1.400[3]$	&& 4.016[3]	&	1.330[3]\\
$^4P_{3/2}$				&$3.973[3]$&	$-1.116[3]$	&& 3.957[3]	&	$-1.117[3]$\\
$^4P_{1/2}$ 				&$8.586[3]$&	$0.0$ 		&& $8.532[3]$ 	& $0.0$	\\\\
$2s2p^4$ $^2D_{5/2}$ 		&$4.668[3]$&	$-2.886[3]$	&& 4.630[3]	&$-2.877[3]$\\
$^2D_{3/2}$				&$-1.627[3]$& 	$-2.021[3]$	&&$-1.598[3]$	&$-2.014[3]$	\\\\
$2s2p^4$ $^2S_{1/2}$ 		&$1.721[4]$&	$0.0$		&&	1.704[4] 	&	0.0 \\\\
$2s2p^4$ $^2P_{3/2}$ 		&$-9.208[2]$& 	$1.402[3]$	&&$-9.124[2]$&	1.392[3]\\
 $^2P_{1/2}$ 		&$2.678[3]$&	$0.0$		&& 2.616[3]	&	0.0 \\\\
$2p^5$ $^2P^o_{3/2}$ 		&$5.684[2]$&	$-1.346[3]$	&& 5.680[2]	& $-1.342[3]$	\\
$^2P^o_{1/2}$				&$4.237[3]$&	$0.0$		&& 4.197[3]	& 0.0 \\
\end{tabular}
\end{table}

\normalsize\bigskip

\end{document}